\documentclass[12pt]{article}

\title{\vspace{20mm} \textbf{Light-cone reduction of \\[1.5mm] Witten's open string field theory} \vspace{10mm} }
\author{\Large Hiroaki Matsunaga \vspace{3mm} } 
\date{
\textit{CEICO, Institute of Physics of the Czech Academy of Sciences, \\[0mm] Na Slovance 2, Prague 8, Czech Republic \\[3mm] Institute of Mathematics of the Czech Academy of Sciences, \\[0mm] Zitna 25, Praha 1, Czech Republic } \\ \vspace{5mm} matsunaga@fzu.cz \vspace{12mm} }

\usepackage[dvips]{graphicx}
\usepackage{color, upgreek, mathrsfs, enumitem}
\usepackage{amsmath,amscd, amssymb, amsfonts, amsthm}

\usepackage[hypertex]{hyperref}
\usepackage{cite}

\makeatletter
  
  \@addtoreset{equation}{section}
\makeatother

\newcommand{\no}{\nonumber\\}

\newcommand{\ld}{ [ \hspace{-0.6mm} [ }
\newcommand{\rd}{ ] \hspace{-0.6mm} ] }
\newcommand{\Ld}{ \big[ \hspace{-1.1mm} \big[ }
\newcommand{\Rd}{ \big] \hspace{-1.1mm} \big] }
\newcommand{\LD}{ \Big[ \hspace{-1.3mm} \Big[ }
\newcommand{\RD}{ \Big] \hspace{-1.3mm} \Big] }
\newcommand{\la}{\big{\langle }}
\newcommand{\ra}{\big{\rangle }}
\newcommand{\La}{\Big{\langle }}
\newcommand{\Ra}{\Big{\rangle }}

\newcommand{\1}{\mbox{1}\hspace{-0.25em}\mbox{l}}

\newcommand{\bM}{\mathbf{M}}

\newcommand{\cA}{\mathcal{A}} 
 
\newcommand{\cC}{\mathcal{C}} 
\newcommand{\cD}{\mathcal{D}}

\newcommand{\cG}{\mathcal{G}}
\newcommand{\cH}{\mathcal{H}}

\newcommand{\cL}{\mathcal{L}} 
 
\newcommand{\cN}{\mathcal{N}}

\newcommand{\cT}{\mathcal{T}} 
\newcommand{\cU}{\mathcal{U}}

\allowdisplaybreaks[3] 

\begin{document}

\maketitle \thispagestyle{empty}\addtocounter{page}{-1} 

\abstract{
We elucidate some exact relations between light-cone and covariant string field theories 
on the basis of the homological perturbation lemma for $A_{\infty }$. 
The covariant string field splits into the light-cone string field and trivial excitations of BRST quartets: 
The latter generates the gauge symmetry and covariance. 
We first show that the reduction of gauge degrees can be performed by applying the lemma, 
which gives a refined version of the no-ghost theorem of covariant strings. 
Then, 
we demonstrate that after the reduction, 
gauge-fixed theory can be regarded as a kind of effective field theory and it provides an exact gauge-fixing procedure taking into account interactions. 
As a result, 
a novel light-cone string field theory is obtained from Witten's open string field theory. 
}

\clearpage

\tableofcontents 

\section{Introduction}

The covariant formulation of string fields enables us to treat a multi-body system of strings, 
which should be a useful tool. 
However, 
except for bosonic open strings, 
covariant string fields based on the minimal world-sheet variables often require impractical treatments \cite{Witten:1985cc, Zwiebach:1992ie}. 
Besides it, 
the light-cone formulation of string fields has long been known \cite{Kaku:1974zz}, 
which is another option. 
Although the obvious Lorentz covariance is lost, 
it gives an independent, 
consistent and easy-to-handle interacting theory. 
It has been applied to various types of researches so far, 
in which peculiar calculation techniques were developed. 

\vspace{2mm}

It may seem surprising, 
but the relation between these covariant and light-cone formulations has remained mysterious. 
They are independently formulated in different ways and we have no dictionary to translate calculations from one side to the other. 
It is important to relate the light-cone and covariant formulations concretely, 
which is our ultimate aim. 
We need to know when the light-cone formulation recovers covariance,\footnote{In this direction, 
there are some interesting investigations. 
See \cite{Hata:1986jd, Kugo:1987rq ,Siegel:1987ku} for example. } 
how the light-cone string field appears in the covariant formulation,
and what the difference is. 
In this paper, 
as a first step, 
we consider the light-cone reduction within the covariant formulation. 

\vspace{2mm}

An analogy with the usual field theory may suggest that the light-cone formulation can be obtained from the covariant formulation via some gauge-fixing---such naive expectation will be correct just for the free theory of string fields. 
The covariant kinetic term indeed reduces to the light-cone kinetic term thanks to the no-ghost theorem of strings proved at the dawn of the covariant formulation \cite{Kato:1982im}. 
We generalize it to the interacting theory and show that a novel light-cone string field theory appears within the covariant formulation, 
from which we cannot extract the old light-cone formulation itself directly. 

\vspace{2mm} 

Note that gauge-fixed theory can be regarded as a kind of effective theory. 
For a given gauge theory $S[\Psi ]$, 
one can obtain a gauge-fixed action $S_{\mathrm{red}} [\psi ]$ by integrating out the gauge degrees $\psi _{g}$ of $\Psi = \psi + \psi _{g} $ as 
\begin{align}
\label{effective} 
\int \cD [\Psi ] e^{-S [\Psi ] } 
= 
\int \cD [\psi , \psi _{g} ] e^{-S [\psi + \psi _{g} ] } 
= ( \mathrm{Vol}_{g} ) \cdot 
\int \cD [\psi  ] e^{-S_{\mathrm{red}} [\psi  ] } 
\, , 
\end{align}
where $\mathrm{Vol}_{g}$ denotes its gauge volume. 
It will give an exact gauge-fixing procedure taking into account interactions. 
The reduced action $S_{\mathrm{red}}[\psi ]$ reproduces the same amplitudes as the original action $S[\Psi ]$. 
As we will show, 
the homological perturbation lemma provides us exact treatment of this formal procedure. 
In particular, 
by applying the lemma for $A_{\infty }$ algebras, 
an $A_{\infty }$ effective field theory is directly obtained from the original $A_{\infty }$ field theory. 
In this paper, 
we construct an action for light-cone string field theory explicitly as the classical part of such an \textit{effective} action for the Witten's theory. 

\vspace{2mm} 

In section 2, 
we review the relation between the BRST operator and the light-cone kinetic operator. 
There exist similarity transformations connecting these. 
In section 3, 
we explain the homological perturbation lemma for $A_{\infty }$. 
We show that the reduction of gauge degrees can be described by applying the lemma, 
which provides a refined version of the no-ghost theorem of covariant strings. 
In section 4, 
we explain the light-cone reduction of interacting theory. 
A novel light-cone string field theory is constructed, 
which has an $A_{\infty }$ action. 
Appendix A is devoted to explaining basic facts of the homological perturbation and its application to similarity transformations. 

\vspace{2mm} 

In this paper, 
we write $\ld A , B \rd $ for the graded commutator of $A$ and $B$, 
\begin{align*}
\Ld \, A \, , \, B \, \Rd \equiv A \, B - (-)^{AB} B \, A \, , 
\end{align*} 
where the upper index of $(-)^{A}$ denotes $A$'s degree. 
The graded commutator will be defined for states, operators or mathematical operations appropriately.

\section{On the kinetic operator}

In this section, 
we briefly review how the light-cone kinetic operator appears in the BRST operator of strings. 
See textbooks or manuscripts treating the no-ghost theorem or BRST cohomology of covariant strings: 
For example, 
section 3 of \cite{Aisaka:2004ga} is pedagogical. 

\vspace{2mm} 

Let us consider bosonic open strings in the flat space-time.   
Recall that the kinetic operator $K^{lc}$ of the light-cone gauge string theory takes the form 
\begin{align}
\label{LC op}
K^{lc} \equiv \frac{1}{2} p^{2} 
+ \sum_{n \geq 1} a^{I}_{-n} a^{I}_{n} - 1 \, , 
\end{align}
where $p_{\mu }$ denotes the momentum zero mode and $a^{I}_{n}$ for $0 < I < 25$ denotes a transverse component of the matter excitation mode $a^{\mu }_{n}$ for $0 \leq \mu \leq 25$. 
The canonical commutation relation of $a^{\mu }_{n}$ is given by $\ld a^{\mu }_{m} , a^{\nu }_{n} \rd = m \eta ^{\mu \nu } \delta _{m+n , 0}$, 
where $\eta ^{\mu \nu }$ is the flat Minkowski metric. 
We introduce the light-cone coordinates for $X^{\mu }$ as follows, 
\begin{align*}
X^{\pm } \equiv \frac{1}{\sqrt{2} } \big{(} X^{0} \pm X^{25} \big{)} \, , 
\end{align*}
where $a^{\pm }_{n} \equiv \frac{1}{\sqrt{2} } ( a^{0}_{n} \pm a^{25}_{n} )$ satisfies $\ld a^{\pm }_{m} , a^{\mp }_{n} \rd = - m \delta _{m+n,0}$ and $\ld a^{\pm }_{m} , a^{\pm }_{n} \rd =0$. 
Likewise, 
we write $p^{\pm } \equiv \frac{1}{2} (p^{0} \pm p^{25} )$ for the light-cone mode of the momentum. 
As we will show, 
a pair of light-cone modes $\{ a^{+}_{n} , a^{-}_{-n} \}$ and $bc$-ghost modes $\{ c_{n} , b_{-n} \}$ gives a BRST quartet. 

\subsection{Light-cone decomposition of BRST}

We explain how the light-cone kinetic operator $K^{lc}$ appears in the BRST operator $Q$. 
Let us consider the ghost-zero-mode decomposition of the BRST operator of open strings 
\begin{align}
\label{BRST op}
Q & \equiv \sum _{n} c_{-n} L_{n} 
- \frac{1}{2} \sum_{n,m} (m-n) \, : c_{-m} c_{-n} b_{m+n} :  
\no 
& =  c_{0} \Big[ L_{0} 
+ \sum_{n \not= 0} n \, : c_{-n} b_{n} : \Big] 
- b_{0} \sum_{n \not=0 } n \, : c_{-n} c_{n} : 
+ \, Q' \, , 
\end{align}
where the symbol $: \hspace{0mm} :$ denotes the normal ordering and $Q'$ consists of the nonzero modes 
\begin{align*} 
Q' & \equiv  
\sum _{n \not=0 } c_{-n} L_{n} 
- \frac{1}{2} \sum_{ \genfrac{}{}{0pt}{1}{n,m \not= 0 }{n+m \not= 0 } } 
(m-n) :c_{-m} c_{-n} b_{n+m} : 
\, .
\end{align*}
The $bc$-ghost modes satisfy the canonical commutation relation $\ld b_{m} , c_{n} \rd = \delta _{m+n,0}$. 
We consider the light-cone decomposition of Virasoro generators via $a^{\pm }_{n} = \frac{1}{\sqrt{2} } (a^{0}_{n} \pm a^{25}_{n} )$. 
The matter Virasoro zero mode can be cast as  
\begin{align*}
L_{0} & = \frac{1}{2} p^{2} + \sum_{n \geq 1} a^{\mu}_{-n} a_{\mu n} - 1
= K^{lc} 
+ \sum_{n \geq 1} \big[ a^{+}_{-n} a^{-}_{n} + a^{-}_{-n} a^{+}_{n} \big]
\, , 
\end{align*}
in which the light-cone kinetic operator, $K^{lc}$ defined by (\ref{LC op}), naturally appears. 
The other Virasoro generators of $X^{\mu }$ are given by 
\begin{align*}
L_{n} = \frac{1}{2} \sum_{k} a^{\mu }_{n-k} a_{\mu  k} 
= - p^{+} a^{-}_{n} 
- \sum_{k \not= 0} a^{+}_{n-k} a^{-}_{k} 
+ \frac{1}{2} \sum_{k} a^{i}_{n-k} a^{i}_{k}
- p^{-} a^{+}_{n} 
\, 
\end{align*}
for $n \not= 0$, 
where $p^{\pm }_{n} \equiv \frac{1}{\sqrt{2}} ( a^{0}_{n} \pm p^{25}_{n} )$. 
Note that the level counting operator $N$ acting on quartets $\{ a^{+}_{n} , c_{n} , b_{-n} , a^{-}_{-n} \} _{n\not= 0}$ naturally appears in the $c_{0}$-part of (\ref{BRST op}): 
\begin{align}
\label{counting op}
N \equiv  - \sum_{n \geq 1} \Big[ a^{+}_{-n} a^{-}_{n} + a^{-}_{-n} a^{+}_{n} - n \big( c_{-n} b_{n} + b_{-n} c_{n} \big) \Big] \, . 
\end{align} 
We find differential operators acting on the quartet for $p^{+} \not= 0$ or $p^{-} \not= 0$, 
\begin{align*}
d\equiv - p^{+} \sum _{n \not= 0} c_{-n} a^{-}_{n} \, , 
\hspace{5mm} 
\bar{d} \equiv -p^{-} \sum_{n \not= 0} c_{-n} a^{+}_{n} \, . 
\end{align*} 
These are nilpotent and have no cohomology, 
which we will see later. 
Since $c_{0} K^{lc}$ is nilpotent and satisfies $\ld c_{0} K^{lc} , d \, \rd = 0$, 
the operator $c_{0} K^{lc} + d$ is also nilpotent. 

\vspace{2mm} 

We find that the nonzero mode part $Q'$ includes these two nilpotent operators and takes the form $Q ' = d + Q_{1} + \bar{d}$ where 
\begin{subequations} 
\begin{align}
\label{Q1}
Q_{1} & \equiv \sum_{n \not= 0} c_{-n} \Big[ 
\frac{1}{2} \sum_{k} a^{i}_{n-k} a^{i}_{k} 
- \sum_{k \not= 0} a^{+}_{n-k} a^{-}_{k} 
\Big] 
- \frac{1}{2} \sum_{ n,m \not= 0 \atop  n+m \not= 0} (m-n) :c_{-m} c_{-n} b_{n+m} : \, . 
\end{align} 
Namely, 
the nilpotent operator $c_{0} K^{lc}$ or $c_{0} K^{lc} + d$ appears in the BRST operator (\ref{BRST op}). 
When a nilpotent operator included in $Q$ has the same cohomology as $Q$, 
the other term has no cohomology and can be regarded as a \textit{perturbation}. 
We define perturbing terms 
\begin{align}
Q_{2} & \equiv - b_{0} \sum_{n \not= 0} n \, : c_{-n} c_{n} : \, , 
\\ 
Q_{3} & \equiv \bar{d} = -p^{-} \sum_{n \not= 0} c_{-n} a^{+}_{n} \, . 
\end{align}
\end{subequations} 
As a result, 
we obtain the light-cone decomposition of the BRST operator 
\begin{align}
\label{lc dec} 
Q = d + c_{0} \Big[ K^{lc} + N \Big] + \sum_{k=1}^{3} Q_{k} \, , 
\end{align}
in which $d$, $c_{0} N$ and $\sum _{k} Q_{k}$ are perturbations connecting $c_{0} K^{lc}$ to $Q$. 
One can use each of them as a perturbation: 
For example, 
$c_{0} N + \sum_{k} Q_{k}$ connects $c_{0} K^{lc} + d$ to $Q$. 
These connections can be understood as maps between nilpotent operators. 
We can find that there exists a similarity transformation $\cU $ between $Q$ and $c_{0} K^{lc} + d$ as follows 
\begin{align}
\label{lc of BRST}
Q  = \cU ^{-1} \big{(} d + c_{0} K^{lc} \big{)} \, \cU \, . 
\end{align}
This is a refined form of the light-cone decomposition of the BRST operator (\ref{lc dec}), 
which is our starting point in this paper. 
The important fact is that $d$ has no cohomology and such a linear map $\cU $ exists and provides (\ref{lc of BRST}). 
It implies that $\cU \, Q = (c_{0} K^{lc} + d ) \, \cU $ provides a morphism of two $A_{\infty }$ algebras preserving its cohomology. 

\vspace{2mm} 

In the rest of this section, 
we first show that $d$ acting on the BRST quartet has no cohomology. 
Then, 
we construct the similarity transformation $\cU $ explicitly with some computations, 
which follows \cite{Aisaka:2004ga}. 
One can construct $\cU $ in a simple manner by applying the homological perturbation lemma, 
which we explain in appendix A.

\subsection{On the BRST quartet} 

In the BRST framework, 
each pair of $\{ a^{+}_{n} , c_{n} ; b_{-n} , a^{-}_{-n} \} _{n \not= 0}$ forms a trivial quartet. 
We suppose $p^{+} \not= 0$, 
which enables us to shift the $bc$-ghost system to the $p^{+} c$ and $\frac{1}{p^{+}} b$ system. 
We write $q_{+} \equiv \{ a^{+}_{n} ,  p^{+} c_{n}  \} _{n \not= 0}$ and $q_{-} \equiv \{ \frac{1}{p^{+}} b_{n} , a^{-}_{n} \} _{n \not= 0}$ for pairs of nonzero modes. 
Because of the canonical commutation relations, 
a differential operator  
\begin{align} 
\label{quartet d}
d \equiv - p^{+} \sum_{n \not= 0} c_{-n} a^{-}_{n} \, 
\end{align} 
acts on the nonzero modes $q = q_{+} \oplus q_{-}$ and is nilpotent $(d )^{2} = 0$. 
Therefore, 
$d$ generates BRST transformation $\delta _{\mathrm{B}}$ satisfying $(\delta _{\mathrm{B}})^{2} = 0$ as follows 
\begin{align*} 
\delta _{\textrm{B}}  \big{(} a^{+}_{n} \big{)} 
\equiv \Ld \, d \, , \, a^{+}_{n} \, \Rd = - n \big{(} p^{+} c_{n} \big{)} \, , 
\hspace{5mm}  
\delta _{\textrm{B}} \Big{(} \frac{1}{p^{+} } b_{-n} \Big{)} 
\equiv \LD \, d \, , \, \frac{1}{p^{+} } b_{-n} \, \RD = a^{-}_{-n} \, . 
\end{align*} 
We call this type of pair of the excitations $q$ and differential $d$ as a \textit{BRST quartet}. 
The BRST quartet has no cohomology, 
which is a well-known fact. 
For each excitation mode, 
the differential $d$ has no cohomology because commutation relations $\ld p^{+} c_{m} , \frac{1}{p^{+}} b_{n} \rd = \delta _{n+m, 0}$ and $\ld a^{\pm }_{m} , a^{\mp }_{n} \rd = - m \delta _{m+n , 0}$ imply the existence of its homotopy contracting operator. 

\vspace{2mm} 

Since $d$ acts on $q_{+}$ and $q_{-}$ separately, 
we can define the quartet splitting operator 
\begin{align*}
S_{\pm } \equiv - \sum_{n \not= 0} : \Big[ \frac{1}{n} a_{-n}^{+} a_{n}^{-} - c_{-n} b_{n} \Big] : \, ,
\end{align*}
which satisfies $S_{\pm } q_{+} = q_{+}$ and $S_{\pm }q_{-} = - q_{-}$. 
One can construct a kind of homotopy contracting operator $\kappa _{\pm }$ satisfying $d \, \kappa _{\pm } + \kappa _{\pm } \, d = S_{\pm } $ as follows 
\begin{align}
\kappa _{\pm } \equiv \frac{1}{p^{+} } \sum_{n \not= 0} \frac{1}{n} a^{+}_{-n} b_{n} \, .  
\end{align}
When $\Phi $ satisfies $\ld S_{\pm } ,  \Phi \rd = n \Phi$ and $\ld d , \Phi \rd = 0$, we find $\Phi = \frac{1}{n} \ld S_{\pm } , \Phi \rd = \frac{1}{n} \ld d , \ld \kappa _{\pm } , \Phi \rd \rd$ for $n \not= 0$. 
Any physical state therefore includes the same numbers of $q_{+}$- and $q_{-}$-excitations. 
Note that operators $d$, $N$, $c_{0}$ and $K^{lc}$ appearing in (\ref{lc of BRST}) have no $S_{\pm }$-excitation: 
\begin{align*} 
\Ld S_{\pm } , d \, \Rd 
= \Ld S_{\pm } , N \Rd 
= \Ld S_{\pm } , c_{0} \Rd 
= \Ld S_{\pm } , K^{lc} \Rd 
= 0 \, .
\end{align*}  

\vspace{2mm}

Likewise, 
we have the level counting operator $N$ acting on the quartet, 
which is defined by (\ref{counting op}). 
We can quickly find another kind of homotopy contracting operator  
\begin{align}
\label{kappa}
\kappa \equiv \frac{1}{p^{+} } \sum_{n \not= 0 } a^{+}_{-n} b_{n} \, . 
\end{align}
It will provide a standard situation of the homological perturbation lemma. 
We find 
\begin{align*}
d \, \kappa + \kappa \, d = N  \, . 
\end{align*}
When $\Phi $ satisfies $\ld N ,  \Phi \rd = n \Phi$ and $\ld d , \Phi \rd = 0$, we find $\Phi = \frac{1}{n} \ld N , \Phi \rd = \frac{1}{n} \ld d , \ld \kappa , \Phi \rd \rd$ for $n \not= 0$. 
Hence, 
the physical modes of $\Phi $ condense on the $N \Phi = 0$ subspace and BRST quartets' excitations have no cohomology. 

\vspace{2mm} 

Let $\alpha $ be an operator satisfying $\ld \alpha , d \rd = 0$. 
When $(\alpha \kappa )^{2} = 0$ holds, 
we find $e^{- \alpha \kappa } d e^{\alpha \kappa } = d + (-)^{\alpha }\alpha N $. 
Since $c_{0}$ is nilpotent and commutes with $d$ and $\kappa $, 
we obtain a similarity transformation between $c_{0} K^{lc} + d$ and $c_{0} [K^{lc} + N ] + d$ as follows 
\begin{align}
\label{U1}
Q - \sum_{k=1}^{3} Q_{k}  
= c_{0} [ K^{lc} + N ] + d 
= e^{c_{0} \kappa } \big{(} c_{0} K^{lc} + d \big{)} \, e^{-c_{0} \kappa }
\, . 
\end{align}
One can obtain the same result by applying the homological perturbation lemma, 
in which $c_{0} N$ is a perturbation. 
See appendix A.

\subsection{On the similarity transformation} 

As we explain, 
an explicit form of the similarity transformation is given by 
\begin{align}
\label{U=U1U2}
\cU \equiv e^{-c_{0} \kappa } \, e^{\ld \kappa _{\pm } , Q_{1} + \frac{1}{2} Q_{2} \rd } \, . 
\end{align}
Since the map $e^{-c_{0} \kappa }$ connects $Q - \sum_{k} Q_{k}$ to  $c_{0} K^{lc} +d$ as (\ref{U1}), 
the map $e^{\ld \kappa _{\pm } , Q_{1} + \frac{1}{2} Q_{2} \rd }$ generates a similarity transformation between $Q$ and $Q- \sum_{k} Q_{k}$. 
We show that the BRST operator $(\ref{lc dec})$ has the following expression 
\begin{align}
\label{U2}
Q & = e^{- \ld \kappa _{\pm } , Q_{1} + \frac{1}{2} Q_{2} \rd } 
\Big{(} Q - \sum_{k=1}^{3} Q_{k} \Big{)} 
\, e^{\ld \kappa _{\pm } , Q_{1} + \frac{1}{2} Q_{2} \rd } \, . 
\end{align}
As the derivation of (\ref{U1}), 
the similarity transformation (\ref{U2}) can be obtained by brute-force calculations or by using the lemma. 
In this section, 
we explain the former on the basis of a pedagogical approach of \cite{Aisaka:2004ga}. 
For the latter approach, see appendix A. 
 
\vspace{2mm} 

Let us consider commutation relations of $Q_{k}$ for $k=1,2$\,: 
While $Q_{1}$ and $Q_{2}$ commutes with $d$ and $N$, 
they have non-trivial $S_{\pm }$-excitations 
\begin{align*}
\Ld d \, , Q_{k} \Rd = 0 \, , \hspace{5mm} 
\Ld N , Q_{k} \Rd = 0 \, , \hspace{5mm} 
\Ld S_{\pm } , Q_{k} \Rd = k \, Q_{k} \, . 
\end{align*} 
Because of $\ld d , \kappa _{\pm } \rd = S_{\pm }$, 
the operator $Q_{k}$ for $k=1,2$ can be cast as follows  
\begin{align*}
Q_{k} = \Ld d , R_{k+1} \Rd \, , \hspace{5mm} 
R_{k+1} \equiv \frac{1}{k} \Ld \kappa _{\pm } , Q_{k} \Rd \, . 
\end{align*}
We thus find that $\cU = e^{-c_{0} \kappa } e^{R_{2} + R_{3}}$ and the right hand side of (\ref{U2}) gives 
\begin{align*}
e^{- R_{2} - R_{3} }
\big{(} d + c_{0} [ K^{lc} + N ]  \big{)} \, 
e^{R_{2} +R_{3} } 
= d + c_{0} [ K^{lc} + N ] + \sum_{k=1}^{2} Q_{k} + \cdots \, . 
\end{align*}
As we see, 
the above ``$\cdots $'' just equals to $Q_{3}$ and it completes our proof of (\ref{U2}). 
We show that $Q_{3}$ takes the following form and the other higher commutators of $R_{2} + R_{3}$ vanish, 
\begin{align*}
Q_{3} = \Ld c_{0} (K^{lc} + N ) , R_{3} \Rd + \frac{1}{2} \Ld \ld d , R_{2} \rd , R_{2} \Rd \, .  
\end{align*} 

\vspace{2mm} 

For this purpose, 
we use the nilpotent relation $Q^{2} =0$ in terms of (\ref{lc dec}). 
Namely, 
we know $(Q)^{2} = (d + c_{0} [K^{lc} + N] + \sum_{k=1}^{3} Q_{k} )^{2} = 0$, 
which provides the following series of identities 
\begin{subequations} 
\begin{align} 
\label{eq.a}
\big{(} d \, \big{)}^{2} & = 0 \, , 
\\
\Ld d , c_{0} (K^{lc} + N ) \Rd & = 0 \, , 
\\ 
\big{(} c_{0} [K^{lc} + N ] \big{)}^{2} + \Ld d , Q_{1} \Rd & = 0 \, , 
\\ 
\Ld c_{0} (K^{lc} + N ) , Q_{1} \Rd + \Ld d , Q_{2} \Rd & = 0 \, , 
\\ \label{eq.e} 
\big{(} Q_{1} \big{)}^{2} + \Ld d , Q_{3} \Rd + \Ld c_{0} (K^{lc} + N ) , Q_{2} \Rd & = 0 \, , 
\\ \label{eq.f}
\Ld c_{0} (K^{lc} + N )  , Q_{3} \Rd + \Ld Q_{1} , Q_{2} \Rd & = 0 \, , 
\\ \label{eq.g}
\big{(} Q_{2} \big{)}^{2} + \Ld Q_{1} , Q_{3} \Rd & = 0 \, , 
\\ 
\Ld Q_{2} , Q_{3} \Rd & = 0 \, , 
\\ 
\big{(} Q_{3} \big{)}^{2} & = 0 \, .
\end{align} 
\end{subequations}   
One can use these relations instead of direct but complicated computations. 
In addition to these, 
we quickly find the following relations from (\ref{Q1}-c), 
\begin{align*}
\Ld c_{0} (K^{lc} + N) , Q_{3} \Rd = 0 \, , \hspace{5mm}
\Ld Q_{1} , Q_{2} \Rd = 0 \, , \hspace{5mm} 
(Q_{2})^{2} = 0 \, , \hspace{5mm}
\Ld Q_{1} , Q_{3} \Rd = 0 \, ,
\end{align*}
which are little stronger than (\ref{eq.f}) and (\ref{eq.g}). 
By using (\ref{eq.e}), 
we obtain  
\begin{align*}
\Ld \ld d , R_{2} \rd , R_{2} \Rd 
= - \Ld \kappa _{\pm } , \frac{1}{2} \ld Q_{1} , Q_{1} \rd \Rd 
= \underbrace{\Ld \kappa _{\pm } , \ld d , Q_{3} \rd \Rd }_{2 Q_{3} } 
+ \underbrace{\Ld \kappa _{\pm } , \ld c_{0} (K^{lc} +N) , Q_{2} \rd \Rd }_{-2 \ld c_{0} (K^{lc} +N) , R_{3} \rd } \,  
\end{align*}
and $Q_{3}$ can be cast as the above. 
Likewise, 
all unwanted terms vanish thanks to (\ref{eq.a}-i). 
The homotopy contracting operator $\kappa _{\pm }$ satisfies 
\begin{align*}
\Ld \kappa _{\pm } , R_{2} \Rd = 0 \, , \hspace{5mm} 
\Ld \kappa _{\pm } , R_{3} \Rd = 0 \, , \hspace{5mm} 
\Ld \kappa _{\pm } , Q_{3} \Rd = 0 \, , 
\end{align*}
and $R_{3}$ satisfies $\ld Q_{1} , R_{3} \rd = 0$ and $\ld Q_{2} , R_{3} \rd = \Ld \ld d , R_{3} \rd , R_{3} \Rd = 0$, 
which gives 
\begin{align*}  
\Ld Q_{2} , R_{2} \Rd & = 2 \underbrace{ \Ld R_{3} , Q_{1} \Rd }_{=0} - \Ld \kappa _{\pm } , \ld \underbrace{Q_{2} }_{\ld d , R_{3} \rd } , Q_{1} \rd \Rd 
= \Ld \ld d , R_{3} \rd , R_{2} \Rd = 0 \, , 
\\ 
\Ld R_{2} , R_{3} \Rd & = \frac{1}{2} \Big[ \Ld \ld  \kappa _{\pm } , R_{2} \rd , Q_{2} \Rd + \Ld \kappa _{\pm } , \ld Q_{2} , R_{2} \rd \Rd \Big] 
= 0 \, . 
\end{align*} 
We obtain the following relations for $i,j =2 , 3$ 
\begin{align*} 
\LD \Big{(} \Ld \ld d , R_{2} \rd , R_{2} \Rd \Big{)} , R_{i} \RD = 0 \, , 
\hspace{5mm} 
\Ld \ld c_{0} (K^{lc} +N) , R_{i} \rd , R_{j} \Rd = 0 \, , 
\end{align*} 
because of $ \Ld \ld c_{0} (K^{lc} + N) , R_{3} \rd , R_{3} \Rd = 0$ and 
\begin{align*}
\Ld c_{0} (K^{lc} + N ) , \underbrace{\ld \kappa _{\pm } , Q_{1} \rd }_{R_{2}} \Rd & = 
\Ld \underbrace{ \ld c_{0} (K^{lc} + N ) , \kappa _{\pm } \rd }_{=0} , Q_{1} \Rd 
- \Ld \kappa _{\pm } , \underbrace{\ld c_{0} (K^{lc} + N) , Q_{1} \rd }_{= 0} \Rd  
 \, .
\end{align*}

\vspace{2mm} 

We would like to give some comments on the consistency of (\ref{U=U1U2}) and $Q_{3}$. 
One can introduce an intermediate operator $Q (t)$ connecting $Q(0) = d + c_{0} [K^{lc} + N]$ to $Q(1) = Q$ and assume that $\cU (t) = e^{t R}$ provides $Q(t) = \cU ^{-1} (t) \, Q(0) \, \cU (t)$. 
There is alternative derivation of $Q_{3}$ by using $Q_{1} = \ld d , R_{2} \rd$ and $Q_{2} = \ld d , R_{3} \rd$. 
We consider 
\begin{align*}
Q(t) \equiv d + c_{0} [K^{lc} + N ] + t ( Q_{1} + Q_{2} ) + Q_{3} (t) \, , 
\end{align*}
where $Q_{3} (t) $ satisfies $Q_{3} (0) = 0$ and $Q_{3} (1) = Q_{3}$. 
We find that the defining equation of $Q_{3}$, 
\begin{align*}
\frac{d}{dt} Q_{3} (t) = \Ld \, c_{0} ( K^{lc} + N ) , \, R_{3} \, \Rd + t \, \Ld \, Q_{1} , \, R_{2} \, \Rd \, , 
\end{align*}
is derived from the following simple differential equation for $Q(t)$ with $R = R_{2} + R_{3}$, 
\begin{align*}
\frac{d}{dt} Q(t) = \Ld \, Q(t) , \, R_{2} + R_{3} \, \Rd \, . 
\end{align*}
Note that $Q_{3}$ and (\ref{U=U1U2}) give solutions to these equations. 
See also appendix A.

\subsection{Physical states} 

We write $\cH _{\mathrm{cov} }$ for the state space of covariant string fields. 
In the Witten theory, 
the state space $\cH _{\mathrm{cov} }$ has gauge degrees and the physical space is given by its BRST cohomology. 
We explain that after the light-cone decomposition, 
physical states can be described by transverse excitation modes on \textit{the Fock vacuum} $| \Omega \rangle $, 
which relates to the conformal or $\mathrm{SL}(2,\mathbb{R} )$ vacuum $| 0 \rangle $ via $| \Omega \rangle = c_{1} | 0 \rangle $. 
While the conformal vacuum $| 0 \rangle $ is defined by $a^{\mu } _{n} | 0 \rangle = c_{n+1} | 0 \rangle = b_{n-2} | 0 \rangle = 0$ for $n >0$ and satisfies $\langle 0 | c_{-1} c_{0} c_{1} | 0 \rangle \not= 0$, 
the Fock vacuum $|\Omega \rangle $ is defined by $a^{\mu } _{n} | \Omega \rangle = c_{n} | \Omega \rangle = b_{n-1} | \Omega \rangle = 0$ for $n > 0$ and satisfies $\langle \Omega | c_{0} | \Omega \rangle \not= 0$. 
The $\mathrm{SL}(2,\mathbb{R} )$ vacuum $| 0 \rangle $ has the BRST-quartet-excitation number $-1$ 
\begin{subequations} 
\begin{align}
( N + 1)  \big{|}  0  \ra = 0 \, ,  
\end{align}
and the Fock vacuum $| \Omega \rangle = c_{1} | 0 \rangle $ has no excitation of BRST quartets 
\begin{align}
N \big{|} \Omega \ra = 0 \, , 
\end{align}
\end{subequations} 
where the quartet-excitation counting operator $N$ is defined by (\ref{counting op}). 
Since the physical states must have no excitation of BRST quartets,\footnote{Note that $d$ does not annihilate the conformal vacuum: $d \, | 0 \rangle = p^{+} \alpha _{-1}^{-} c_{1} | 0 \rangle \not= 0$ unlike $d \, | \Omega \rangle = 0$. } 
the physical space $\cH _{\mathrm{lc}}$ is given by 
\begin{align*} 
\cH _{\mathrm{lc} } \equiv \mathrm{Span} \Big{(} \, a^{I_{1}}_{-k_{1} } \cdots a^{I_{n} }_{-k_{n} } \big{|} \Omega \ra \, \Big{|} \, 1 \leq k_{1} , \dots k_{n} \, , \hspace{2mm} 0 < I_{1} , \cdots I_{n} < 25 \, \Big{)} \, . 
\end{align*}
We write $\Pi $ for the projection onto the physical space $\cH _{\mathrm{lc}}$ from the state space $\cH _{\mathrm{cov}}$ of covariant string fields, 
namely, $\Pi : \cH _{\mathrm{cov} } \rightarrow \cH _{\mathrm{lc}}$\,. 
Note that $\Pi $ also gives the projector onto $\mathrm{Ker}[N]$, 
the kernel of the BRST-quartet-excitation counting operator $N$. 
Since the operator $N$ has some non-zero value $n$ on $(1 - \Pi )$, 
one can define the operator $\frac{1}{N}$ that gives $n^{-1}$ on $(1- \Pi )$. 
We find $\frac{1}{N} (1 - \Pi ) \Psi = n^{-1} (1 - \Pi ) \Psi $ for any state $\Psi \in  \cH _{\mathrm{cov} }$ satisfying $ N \Psi = n \Psi$ with $n \in \mathbb{R}$ because of the projecting property $(1 - \Pi ) \, \Pi \Psi = 0$. 
We can define a homotopy contracting operator $h$ satisfying $(h)^{2} = 0$ and $h \, \Pi = \Pi \, h = 0$ by 
\begin{align}
\label{lc h}
h \equiv \frac{1}{N p^{+}} \sum_{n\not= 0} a^{+}_{-n} b_{n} \big{(} 1 - \Pi \big{)} \, . 
\end{align} 
Except for the kernel of $N$, 
this $h$ gives an inverse of the BRST differential $d$ as follows 
\begin{align*}
d \, h + h \, d = 1 - \Pi  \, . 
\end{align*} 
As we will see, 
this Hodge type decomposition of the unit enables us to find the light-cone theory. 
While the Fock vacuum $|\Omega \rangle = \Pi \, | \Omega \rangle $ is $h$-closed $h \, | \Omega \rangle = 0$ because of the projection, 
the $\mathrm{SL}(2,\mathbb{R} )$ vacuum $| 0 \rangle = (1 - \Pi ) | 0 \rangle $ is $h$-exact: 
\begin{align*}
\big{|} 0 \ra 
= 
h \, \Big{(} p^{+} a^{-}_{-1} \big{|} \Omega \ra \Big{)}  
\, . 
\end{align*}
The existence of a homotopy contracting operator implies that its cohomology is empty. 
Hence, 
while any physical state belongs to $\cH _{\mathrm{lc} } = \Pi \, \cH _{\mathrm{cov} }$, 
all trivial excitation modes consist of BRST quartet excitations and must belong to $( 1 - \Pi )\, \cH _{\mathrm{cov} }$. 

\vspace{2mm} 

We write $\Psi \in \cH _{\mathrm{cov}}$ for a covariant open string field. 
The kinetic term $S_{2} [\Psi ]$ of the Witten theory is invariant under the gauge transformation, 
\begin{align*}
S_{2} [\Psi ] = \frac{1}{2} \la \Psi , \, Q \, \Psi \ra \, ,  
\hspace{8mm} 
\delta \Psi = Q \, \Lambda \, , 
\end{align*}
where $\Lambda \in \cH _{\mathrm{cov} }$ denotes a gauge parameter field. 
Let us consider the following linear field redefinition 
\begin{align}
\label{redef} 
\Psi _{\mathrm{cov}} \equiv \cU \, \Psi \, . 
\end{align} 
The covariant string field splits into physical and gauge degrees $\Psi _{\mathrm{cov} } \in \cH _{\mathrm{lc} } \oplus ( \1 - \Pi ) \, \cH _{\mathrm{cov} }$. 
It changes the kinetic term and gauge transformation as follows, 
\begin{align}
\label{split action}
S_{2} [\Psi _{\mathrm{cov} } ] = \frac{1}{2} \la \Psi _{\mathrm{cov} } , \, \big{(} c_{0} K^{lc} + d \, \big{)}  \Psi _{\mathrm{cov} } \ra \, , 
\hspace{8mm} 
\delta \Psi _{\mathrm{cov} } = \big{(} c_{0} K^{lc} + d \, \big{)} \Lambda _{\mathrm{cov} } \, , 
\end{align} 
where $\Lambda _{\mathrm{cov}} \equiv \cU \, \Lambda  \in (\1 - \Pi ) \, \cH _{\mathrm{cov} }$ denotes the redefined gauge parameter field. 
Note that $\delta \Psi _{\mathrm{cov} } \in (\1 - \Pi ) \, \cH _{\mathrm{cov} }$. 
If we perform the gauge fixing\footnote{Here, 
the path integral over $(1-\Pi ) \Psi _{\mathrm{cov} }$ is performed. 
Note that $h \, \Psi _{\mathrm{cov}} = 0$ is admissible as a gauge-fixing condition in a usual perturbative field theory because the gauge transformation reduces $\Psi _{\mathrm{cov} }$ to $(\Pi + h d ) \Psi _{\mathrm{cov} }$. 
Then, 
we can get (\ref{lc gauge fixing}) by using the on-shell conditions of gauge and unphysical modes $(1 - \Pi ) \Psi _{\mathrm{cov} } $. } in the sense of (\ref{effective}) by 
\begin{align} 
\label{lc gauge fixing}
N \, \Psi _{\mathrm{cov} } = 0 \, , 
\end{align}
all of the gauge degrees are removed from the covariant string field $\Psi _{\mathrm{cov} }$ because (\ref{lc gauge fixing}) prohibits any excitation of BRST quartets. 
A gauge fixed action is given by 
\begin{align}
\label{gauge fixed free}
S_{2} [ \Psi _{\mathrm{cov} } ] =  \frac{1}{2} \la \Psi _{\mathrm{cov} } , \,  c_{0}  K^{lc} \, \Psi _{\mathrm{cov} } \ra  \, . 
\end{align}
Note that $d \, \Psi _{\mathrm{cov} } = 0$ trivially holds because of (\ref{lc gauge fixing}).  
It just equals to the kinetic term of the old light-cone formulation, 
which we explain in section 3 and 4.

\section{Homological perturbation $\& $ Gauge degrees}

In this section, 
we explain the decoupling mechanism of gauge degrees based on the homological perturbation, 
a powerful mathematical lemma. 
As we will see, 
it gives a refined version of the old no-ghost theorem of covariant strings 
and an exact procedure of partial gauge fixing. 
We elucidate some relations between covariant and light-cone string fields. 
See appendix A for a brief review of the lemma and some application. 
For more rigorous or detailed treatment, 
consult mathematical manuscripts, 
such as \cite{Crainic, Berglund, Vallette}. 
Pedagogical reviews are in \cite{Henneaux, Aisaka:2002sd}. 
See \cite{Kajiura:2003ax, Konopka:2015tta, Erler:2016rxg} for other application to string field theory. 

\subsection{Homological perturbation lemma}

Let $Q$ and $q$ be differentials acting on $\cH $ and $\cL $ respectively. 
We write $\pi$ and $\iota $ for morphisms of two complexes $(\cH ,Q)$ and $(\cL , q)$ preserving its cohomology, 
which satisfy $\pi \, Q = q \, \pi $ and $\iota \, q = Q \,\iota$. 
When a homotopy contracting operator $H$ between $\1 _{\cH }$ and $\iota \, \pi$ exists and $H$ satisfies $Q \, H + H \, Q  = \1 _{\cH } - \iota \, \pi $, it is called as a standard situation: 
\begin{subequations} 
\begin{align}
\label{SS}
H \, 
\rotatebox{-70}{$\circlearrowright $} \, 
\big{(} \cH , Q \big{)} 
\hspace{3mm} 
\overset{\pi }{\underset{\iota }{
\scalebox{2}[1]{$\rightleftarrows $}
}}
\hspace{3mm} 
\big{(} \cL , q \big{)} 
\hspace{7mm}
\mathrm{with} 
\hspace{5mm} 
 \1 _{\cH } - \iota \, \pi = Q \, H + H \, Q  \, . 
\end{align}

A perturbation $\Delta $ of a given standard situation (\ref{SS}) is a map acting on $\cH $ which has the same degree as $Q$ and satisfies $(Q + \Delta )^{2} =0$. 
We assume that $\frac{1}{1 + \Delta H} = \sum (-\Delta H)^{n}$ and $\frac{1}{1+ H \Delta } = \sum ( - H \Delta )^{n}$ are definable. 
Let us introduce a useful operator $A = \Delta  - \Delta H \Delta + \cdots  $ defined by 
\begin{align*}
A \equiv \Delta \sum_{n=0} ( -H \Delta )^{n} = \sum_{n=0} ( -\Delta H )^{n} \, \Delta \, , 
\end{align*} 
which satisfies $A (H \Delta ) = (\Delta H) A = \Delta -A$, $\frac{1}{1+\Delta H} = 1 - A H$ and $\frac{1}{1+ H \Delta } = 1- H A$ by definition. 
Then, 
there exist the perturbed data which also give a standard situation, 
\begin{align}
\label{PD}
H_{\Delta } \, 
\rotatebox{-70}{$\circlearrowright $} \, 
\big{(} \cH , Q_{\Delta }  \big{)} 
\hspace{3mm} 
\overset{\pi _{\Delta }}{\underset{\iota _{\Delta }}{
\scalebox{2}[1]{$\rightleftarrows $}
}}
\hspace{3mm} 
\big{(} \cL , q_{\Delta } \big{)} 
\hspace{7mm}
\mathrm{with} 
\hspace{5mm} 
 \1 _{\cH } - \iota _{\Delta } \, \pi _{\Delta } = Q_{\Delta } \, H_{\Delta } + H_{\Delta } \, Q_{\Delta } \, , 
\end{align}
\end{subequations} 
which is the homological perturbation lemma. 
See appendix A for details. 
In particular, 
the lemma also provides an explicit constructing procedure of the perturbed data. 
The perturbed complexes $(\cH , Q_{\Delta })$ and $(\cL , q_{\Delta })$ are given by nilpotent operators 
\begin{subequations} 
\begin{align}
Q_{\Delta } & \equiv Q + \Delta \, , 
\\ 
q_{\Delta } & \equiv q + \pi \, A \, \iota \, . 
\end{align}
The homological perturbation lemma states that the perturbed operator 
\begin{align}
H_{\Delta } & \equiv H - H \, A \, H \,  
\end{align}
is just a homotopy contracting operator between $\1 _{\cH }$ and $\iota _{\Delta } \, \pi _{\Delta }$ where the perturbed projection $\pi _{\Delta }$ and the perturbed injection $\iota _{\Delta }$ are defined by 
\begin{align}
\pi _{\Delta } & \equiv \pi - \pi \, A \, H \, , 
\\ 
\iota _{\Delta } & \equiv \iota - H \, A \, \iota \, . 
\end{align}
\end{subequations} 
It provides recipes of $q_{\Delta }$ satisfying $(q_{\Delta } )^{2} = 0$ and $H_{\Delta }$ satisfying $\ld Q_{\Delta } , H_{\Delta } \rd = \1 _{\cH } - \iota _{\Delta } \pi _{\Delta }$. 
A proof is in appendix A. 
In the rest of this subsection, 
we give several mathematical definitions for simplicity. 
Afterwords, 
we apply the lemma to the covariant string. 

\subsubsection*{Contractions}

Although the condition (\ref{SS}) may be enough for the lemma, 
it is useful to consider more restricted cases in order to apply the lemma to string field theory. 
A deformation retract is a standard situation having the additional property 
\begin{subequations} 
\begin{align} 
\pi \, \iota = \1 _{\cL} \, . 
\end{align}
When the initial data give a deformation retract, then the perturbed data also give a deformation retract if and only if 
\begin{align*}
\pi \Big[ - A H^{2} A + A H + H A \Big] \iota = 0 \, . 
\end{align*} 
A strong deformation retract, 
or a contraction, 
is a deformation retract satisfying the annihilation properties 
\begin{align}
(H)^{2} = 0 \, , \hspace{5mm } 
H \, \iota = 0 \, , \hspace{5mm } 
\pi \, H = 0 \, .  
\end{align}
\end{subequations} 
When the initial data give a strong deformation retract, then the perturbed data also give a strong deformation retract. 
Note that replacing $H$ by $H (QH+HQ)$ realizes $H \iota = 0$, replacing $h$ by $(QH+HQ)H$ realizes $\pi H = 0$, and replacing $H$ by $HQH$ realizes $(H)^{2} = 0$. 
But it complicates the explicit forms of formulae. 

\vspace{2mm} 

We write $(\cH \overset{\pi }{\underset{\iota }{\scalebox{1.5}[1]{$\rightleftarrows $}}} \cL , H )$ for a contraction for brevity. 
By defining $H \circ K \equiv H + \iota K \pi $, 
we can define the composition of contractions as follows 
\begin{align}
\big{(} \cH \overset{p \pi }{\underset{\iota i }{\scalebox{2}[1]{$\rightleftarrows $}}} \cL  , H + \iota K \pi \big{)}
\equiv
\big{(} \cH \overset{\pi }{\underset{\iota }{\scalebox{2}[1]{$\rightleftarrows $}}} \cN  , H \big{)}
\circ 
\big{(} \cN \overset{p}{\underset{i }{\scalebox{2}[1]{$\rightleftarrows $}}} \cL  , K \big{)} \, . 
\end{align} 
The perturbation lemma is compatible with this composition. 
For example, for a given perturbation $\Delta$, the relation $(\iota i)_{\Delta } = \iota _{\Delta } i _{\Delta }$ holds.

\subsubsection*{Morphism of contractions}

When some theories satisfy the contraction condition, 
a morphism between them may provide us new insights into our understanding of the relation between these theories. 
A morphism of contractions is a morphism of differential graded algebras $\cU : (\cH _{1} , Q_{1}) \rightarrow (\cH _{2} , Q_{2})$ satisfying $\cU \, H_{1} = H_{2} \, \cU $, 
for which we write 
\begin{align}
\label{morphism of contractions}
\cU \, : \, 
\big{(} \cH _{1}\overset{\pi _{1}}{\underset{\iota _{1}}{\scalebox{2}[1]{$\rightleftarrows $}}} \cL _{1} , H_{1} \big{)}
\hspace{1mm}
\longrightarrow 
\hspace{1mm} 
\big{(} \cH _{2} \overset{\pi _{2}}{\underset{\iota _{2}}{\scalebox{2}[1]{$\rightleftarrows $}}} \cL _{2} , H_{2} \big{)} \, . 
\end{align}
Then, we find that $\tilde{\cU } \equiv \pi _{2} \, \cU \iota _{1}$ gives a morphism $\tilde{\cU } : \cL _{1} \rightarrow \cL _{2}$ satisfying $\iota _{2} \, \tilde{\cU } = \cU \, \iota _{1}$ and $\tilde{\cU } \, \pi _{1} = \pi _{2} \, \cU$. 
One would be able to consider the lemma as a useful tool to construct such a morphism explicitly. 
As we show in appendix A, 
we can construct the similarity transformation (\ref{lc of BRST}), (\ref{U1}) or (\ref{U2}) by using it.

\subsection{Reduction of gauge degrees} 

The homological perturbation lemma is a useful tool for describing the reduction of gauge degrees. 
As application, 
we explain that it provides a refined version of the no-ghost theorem of string theory. 
Let $k$ be positive integer $k > 0$. 
For the $k$-th BRST quartet $( a^{+}_{-k} , c_{-k} ; b_{k} , a^{-}_{k} )$, 
an excitation counting operator $N_{k}$ is given by 
\begin{subequations} 
\begin{align}
N_{k} \equiv a^{+}_{-k} a^{-}_{k} - k \, c_{-k} b_{k} \, . 
\end{align}
Likewise, 
for the $(-k)$-th BRST quartet $( a^{+}_{k} , c_{k} ; b_{-k} , a^{-}_{-k} )$, 
we define $N_{-k}$ as follows 
\begin{align}
N_{-k} \equiv a^{-}_{-k} a^{+}_{k} - k \, b_{-k} c_{k} \, . 
\end{align}
\end{subequations} 
Note that the $n$-th BRST quartet has no cohomology and gives a contractible situation with 
\begin{align}
\label{d and h} 
d _{n}  
\equiv - p^{+} c_{-n} a^{-}_{n} \, , 
\hspace{5mm} 
h_{n}  
\equiv \frac{1}{p^{+} N_{n} } a^{+}_{-n} b_{n} \big{(}  \1 - \Pi _{n} \big{)} \, , 
\end{align}
where $\Pi _{n}$ denotes the projector on the subspace without the $n$-th quartet excitations. 
These operators satisfy $(\Pi _{n} )^{2} = \Pi _{n}$, $h_{n} \Pi _{n} = \Pi _{n} = 0$, $d _{n} \Pi _{n} = \Pi _{n} d_{n} = 0$ and 
\begin{align*} 
d_{n} \, h_{n} + h_{n} \, d_{n} = \1 - \Pi _{n} \, .  
\end{align*} 
We find $d = \sum_{n\not= 0} d_{n}$, $h = \sum_{n\not= 0} h_{n}$ and $N = \sum_{k>0} (N_{k} + N_{-k} )$ respectively. 
Let us introduce the $n$-reduced state space 
\begin{align*}
\cH ^{[n]} & \equiv \left[ \prod_{k= -n}^{n} \Pi _{k} \right] \cH _{\mathrm{cov} }
\no 
& = \mathrm{Span} \Big{(} a_{-m_{a}}^{\pm } \cdots a_{-m_{1}}^{\pm } 
b_{-m_{b}} \cdots b_{-m_{1}} 
c_{-m_{c} } \cdots c_{-m_{1}} 
| \, \mathrm{lc} \, \rangle \, \Big{|} \, |n| < | m | \, , \, | \mathrm{lc} \rangle \in \cH _{\mathrm{lc }} \Big{)} \, . 
\end{align*}
While the zeroth space $\cH ^{[0]}$ is just the state space $\cH _{\mathrm{cov}}$ of the covariant formulation, 
the $\infty $-reduced space $\cH ^{[\infty ]}$ is the state space $\cH _{\mathrm{lc}}$ of the light-cone formulation. 
There is a sequence of the reduced state spaces 
\begin{align*}
\cH _{\mathrm{lc} } \equiv \cH ^{[\infty ]} 
\,\, \subset  \,\, \cdots \,\, \subset \,\, 
\cH ^{[n]}  \,\, \subset \,\, \cdots \,\, \subset \,\, 
\cH ^{[1]} \,\, \subset \,\, 
\cH ^{[0]} \equiv \cH _{\mathrm{cov} } \, . 
\end{align*}

By taking $Q = d_{n}$, $H = h_{n}$ and $\cL = \Pi _{n} \cH$ with natural injection and projection, 
the homological perturbation lemma describes the process removing the $n$-th BRST quartet from the theory. 
One can apply this procedure to each BRST quartet successively and finally obtain the theory that consists of physical degrees only. 
It gives reduction of gauge degrees. 
We write $\Psi ^{[n]}$ for a string field living in $\cH ^{[n]}$ and $\Lambda ^{[n]} \in \cH ^{[n]}$ for its gauge parameter. 
The equations of motion and gauge transformation are reduced as follows 
\begin{align*}
&
\begin{cases} 
\big{(} c_{0} K^{lc} + d \big{)} \Psi _{\mathrm{cov} } = 0 
\\ 
\delta \Psi _{\mathrm{cov} } = \big{(} c_{0} K^{lc} + d \big{)} \Lambda _{\mathrm{cov} }
\end{cases} 
\longrightarrow \hspace{2mm} 
\begin{cases} 
\big{(} c_{0} K^{lc} + \sum_{|n| > 1} d_{n} \big{)} \Psi ^{[1]} = 0 
\\ 
\delta \Psi ^{[1]} = \big{(} c_{0} K^{lc} + \sum_{|n| > 1} d_{n} \big{)} \Lambda ^{[1]}
\end{cases} 
\longrightarrow \hspace{2mm} \cdots 
\no & \hspace{5mm} 
\cdots \hspace{2mm} 
\longrightarrow \hspace{2mm} 
\begin{cases} 
\big{(} c_{0} K^{lc} + \sum_{|n| > m} d_{n} \big{)} \Psi ^{[m]} = 0 
\\
\delta \Psi ^{[m]} = \big{(} c_{0} K^{lc} + \sum_{|n| > m} d_{n} \big{)} \Lambda ^{[m]}
\end{cases} 
\longrightarrow \hspace{2mm} 
\cdots 
\hspace{2mm} \longrightarrow \hspace{2mm} 
\begin{cases} 
c_{0} K^{lc} \, \Psi _{\mathrm{lc} } = 0 
\\ 
\delta \Psi _{\mathrm{lc} } = 0  
\end{cases} 
\end{align*}
The no-ghost theorem is equivalent to the perturbed data obtained by setting $Q = d$, $H =h$ and $\cL = \cH _{\mathrm{lc}}$ in the initial data and by taking $\Delta = c_{0} K^{lc}$ as a perturbation. 

\vspace{3mm} 

Let us consider the string field redefinition $\Psi _{\mathrm{cov} } \equiv \cU \, \Psi $ given by (\ref{redef}), which enables us to get the free action in the split form (\ref{split action}). 
We can fix the gauge symmetry generated by the $n$-th BRST quartet by imposing a partially-gauge-fixing condition 
\begin{align}
N_{k} \, \Psi _{\mathrm{cov} } = 0 \, . 
\end{align}
The state $\Psi ^{[n]} \in \cH ^{[n]}$ equals to $\Psi _{\mathrm{cov} }$ satisfying the set of gauge conditions $N_{k} \Psi _{\mathrm{cov} } = 0$ for $|k| \leq | n |$. 
Because of $d_{k} \Psi ^{[n]} = 0$ for $|k| \leq |n|$, 
a partially-gauge-fixed action is given by 
\begin{align}
S_{2} [\Psi ^{[n]} ] = 
\frac{1}{2} \la \Psi ^{[n]} , \, \big{(} c_{0} K^{lc} + \sum_{|k|>n} d_{k} \big{)} \Psi ^{[n]} \ra  \, . 
\end{align}
It has the residual gauge invariance $\delta \Psi ^{[n]} = ( c_{0} K^{lc} + \sum_{|k|>n} d_{k} ) \Lambda ^{[n]}$. 
We can continue this partially-gauge-fixing procedure and finally obtain the light-cone kinetic term (\ref{gauge fixed free}). 
There is no gauge degree in (\ref{gauge fixed free}), 
for $\Psi _{\mathrm{lc}} \in \cH _{\mathrm{lc} }$ is a state carrying ghost number $1$. 
Since any BRST quartet's excitation on the Fock vacuum $| \Omega \rangle $ is prohibited in $\cH _{\mathrm{lc}}$, 
there is no state carrting ghost number $0$ in $\cH _{\mathrm{lc} }$ and thus $(c_{0} K^{lc} )^{2} = 0$ generates no gauge transformation.

\subsection{Homological perturbation for $A_{\infty }$}

The homological perturbation lemma goes well for coalgebras and operadic algebras, 
such as $A_{\infty }$ or $L_{\infty }$. 
It enables us to obtain an off-shell interacting version of the gauge decoupling mechanism. 
We give a brief review of transferring the lemma to $A_{\infty }$. 
Afterwords, 
we apply it to constructing the minimal model of $A_{\infty }$, 
which gives the $S$-matrix.

\subsubsection*{Coalgebra contraction} 

We introduce a \textit{contraction for coalgebras} by using the tensor product of (algebra) contractions. 
For given contractions $\big{\{ } (\cH _{n} \overset{\pi _{n} }{\underset{\iota _{n} }{\scalebox{2}[1]{$\rightleftarrows $}}} \cL _{n} , H_{n} ) \big{\} }_{n}$, 
we define a homotopy contracting operator $H_{1} \ast H_{2}$ acting on the tensor $\cH _{1} \otimes \cH _{2}$ by  
\begin{align}
H_{1} \ast H_{2} \equiv H_{1} \otimes \iota _{2} \pi _{2} + \1 \otimes H_{2} 
\, . 
\end{align}
It enables us to define a tensor product of contractions as follows 
\begin{align*}
\big{(} \cH _{1} \otimes \cH _{2} \overset{\pi _{1} \otimes \pi _{2} }{\underset{\iota _{1} \otimes \iota _{2} }{\scalebox{4}[1]{$\rightleftarrows $}}} \cL _{1} \otimes \cL _{2} \, , \, H_{1} \ast H_{2} \big{)} \, . 
\end{align*} 
A tensor product of contractions also gives a contraction. 
Likewise, using 
\begin{align*} 
T^{n} H \equiv \sum_{i=1}^{n} \1 ^{\otimes i-1} \otimes H \otimes (\iota \pi )^{\otimes n-i }
\, , 
\end{align*}
we define the $n$-fold tensor product of contractions 
\begin{align*}
\bigotimes _{n=1}^{n} 
\big{(} \cH  \overset{\pi  }{\underset{\iota  }{\scalebox{2}[1]{$\rightleftarrows $}}}  \cL  \, , \, H \big{)} 
\equiv 
\big{(} \cH^{\otimes n} \overset{ \pi ^{\otimes n} }{\underset{\otimes \iota ^{\otimes n} }{\scalebox{3}[1]{$\rightleftarrows $}}} \cL ^{\otimes n}  \, , \, T^{n} H \big{)} \, . 
\end{align*}

Let us consider a \textit{coalgebra} $\cC$ with a coproduct $\Delta ' : \cC \rightarrow \cC \otimes ' \cC $ defined by the tensor product.\footnote{We write a prime on the tensor product $\otimes '$ to conceptually distinguish from the tensor product $\otimes $ defining the tensor algebra, 
although these are practically the same in our computations. 
For example, 
we would like to regard $1 \in \mathbb{C}$, 
$1 \otimes 1 \in \cT ( \cH )$ and $1 \otimes ' 1 \in \cT ( \cH ) \otimes ' \cT (\cH )$ to clarify mathematical manipulations or definitions, 
although we may use $1 = 1 \otimes 1 = 1 \otimes ' 1$ in practice. 
} 
We assume that $\cC $ has a contraction $(\cC \, \overset{\pi }{\underset{\iota }{\scalebox{1.5}[1]{$\rightleftarrows $}}} \, \cL _{\, \cC } , H)$. 
When the coproduct 
\begin{align}
\Delta ' \, : \, 
\big{(} \, \cC \, \overset{\pi }{\underset{\iota }{\scalebox{2}[1]{$\rightleftarrows $}}} \, \cL _{\, \cC }  \, , \, H \, \big{)} 
\hspace{1mm} 
\longrightarrow 
\hspace{1mm} 
\big{(} \, \cC  \otimes ' \cC \, \overset{\pi \otimes ' \pi }{\underset{\iota \otimes ' \iota }{\scalebox{3}[1]{$\rightleftarrows $}}} \, \cL _{\, \cC } \otimes ' \cL _{\, \cC } \, , \, H \ast H \, \big{)} 
\end{align}
is a morphism of contractions (\ref{morphism of contractions}), 
it is called as a \textit{coalgebra contraction}. 
Conversely, 
a contraction $(\cH \overset{\pi }{\underset{\iota }{\scalebox{1.5}[1]{$\rightleftarrows $}}} \cL , H)$ becomes a coalgebra contraction when $\pi $ and $\iota $ are morphisms of differential graded \textit{coalgebras} and $H$ satisfies 
\begin{align*}
\big{(} \1 \otimes ' H + H \otimes ' \iota \pi \big{)} \, \Delta ' = \Delta ' \, H \, . 
\end{align*}

\subsubsection*{Tensor trick $\& $ Thick map} 

Let us introduce a contraction for tensor coalgebras. 
For a given contraction $(\cH \overset{\pi }{\underset{\iota }{\scalebox{1.5}[1]{$\rightleftarrows $}}} \cL , H)$ of differential graded algebras, we can consider its tensor coalgebra 
\begin{align*}
\cT ( \cH ) \equiv \bigoplus _{n=0}^{\infty } \cH ^{\otimes n} 
\end{align*}
with the coproduct $\Delta '$ of $\phi _{1} \otimes \cdots \otimes \phi _{n} \in \cH ^{\otimes n}$ defined by 
\begin{align*}
\Delta ' ( \phi _{1} \otimes \cdots \otimes \phi _{n} ) = \sum_{i=1}^{n-1} ( \phi _{1} \otimes \cdots \otimes \phi _{i} ) \otimes ' (\phi _{i+1} \otimes \cdots \otimes \phi _{n} ) \, \Delta ' \, .
\end{align*}
The differential $Q$ acting on $\cH $ is lifted to the differential $\boldsymbol{Q}_{\cT }$ acting on $\cT (\cH )$ via 
\begin{subequations} 
\begin{align} 
\label{cod Q}
\Delta ' \, \boldsymbol{Q}_{\cT } & = \big{(} \boldsymbol{Q}_{\cT } \otimes ' \1 _{\cT } + \1 _{\cT } \otimes ' \boldsymbol{Q}_{\cT } \big{)} \, \Delta ' \, , 
\end{align}
where $\1 _{\cT }$ denotes a unit of the tensor coalgebra $\cT (\cH )$. 

\vspace{2mm} 

We define natural extensions of injection $\iota $, projection $\pi $, their composition $\Pi = \iota \circ \pi $, and contracting homotopy $H$ acting on the tensor coalgebra $\cT (\cH )$ as follows 
\begin{align*}
\boldsymbol{\iota }_{\cT } \equiv \sum \iota ^{\otimes n} \, , \hspace{5mm} 
\boldsymbol{\pi }_{\cT } \equiv \sum \pi ^{\otimes n} \, , \hspace{5mm} 
\boldsymbol{\Pi }_{\cT } \equiv 
\sum ( \iota \circ \pi )^{\otimes n}  \, , \hspace{5mm} 
\boldsymbol{h}_{\cT } \equiv \sum T ^{n} H  \, . 
\end{align*}
Note that $\boldsymbol{\iota }_{\cT }$, $\boldsymbol{\pi }_{\cT }$, and $(\boldsymbol{\iota \pi })_{\cT } \equiv \boldsymbol{\Pi }_{\cT } = \boldsymbol{\iota }_{\cT } \circ \boldsymbol{\pi }_{\cT }$ are morphisms of tensor coalgebras  
\begin{align} 
\label{coh iota}
\Delta ' \, \boldsymbol{\iota }_{\cT } & = \big{(} \boldsymbol{\iota }_{\cT } \otimes ' \boldsymbol{\iota }_{\cT } \big{)} \, \Delta ' \, , 
\\ \label{coh pi} 
\Delta ' \, \boldsymbol{\pi }_{\cT } & = \big{(} \boldsymbol{\pi }_{\cT } \otimes ' \boldsymbol{\pi }_{\cT } \big{)} \, \Delta ' \, .
\end{align}
Using these operations, 
we can define a contraction for tensor coalgebras 
\begin{align*}
\big{(} \, \cT (\cH ) \overset{ \boldsymbol{\pi }_{\cT } }{\underset{\boldsymbol{\iota }_{\cT } }{\scalebox{3}[1]{$\rightleftarrows $}}} \cT ( \cL )  \, , \, \boldsymbol{h}_{\cT } \, \big{)} \, . 
\end{align*} 
It gives a coalgebra contraction because of   
\begin{align} 
\label{coc h} 
\Delta ' \, \boldsymbol{h}_{\cT } & = \big{(} \1 _{\cT } \otimes ' \boldsymbol{h}_{\cT } + \boldsymbol{h}_{\cT } \otimes ' (\boldsymbol{\iota \pi } )_{\cT }  \big{)} \, \Delta ' \, .
\end{align}
\end{subequations} 
We omit the lower $\cT$-index for simplicity in the rest. 
Let us introduce more useful notation. 
A thick map $\boldsymbol{f} : \cT (\cH ) \rightarrow \cT (\cN )$ is a sequence of maps of the same degree, 
\begin{align}
\label{thick} 
\boldsymbol{f} \equiv \big{\{} \boldsymbol{f}_{n} : \cH ^{\otimes n} \rightarrow \cN ^{\otimes n} \big{\} }_{n\geq 0} \, . 
\end{align}
One can lift any coderivation, cohomomorphism or homotopy contraction to a thick map in a natural and trivial way. 
See \cite{Berglund} for further details. 
Clearly, it is compatible with differentials, compositions, and $\mathbb{C}$-linear structure 
\begin{align*}
d ( \boldsymbol{f} )_{n} & = d_{\cN ^{\otimes n} } \boldsymbol{f}_{n} - (-)^{\boldsymbol{f}} \boldsymbol{f}_{n} d_{\cH ^{\otimes n} } \, , 
\\  
(\boldsymbol{f} \circ \boldsymbol{g} )_{n} & = \boldsymbol{f}_{n} \circ \boldsymbol{g}_{n} \, , 
\\  
( a \, \boldsymbol{f} + b \, \boldsymbol{g})_{n} & = a \, \boldsymbol{f}_{n} + b \, \boldsymbol{g}_{n} \, . 
\end{align*}
A thick map $\boldsymbol{f}$ is a morphism if $\boldsymbol{f}_{p+q}=\boldsymbol{f}_{p} \otimes ' \boldsymbol{f}_{q}$ for any $p,q \geq 0$. 
Let $\boldsymbol{l}$ and $\boldsymbol{r}$ be morphisms. 
A thick map $\boldsymbol{d}$ is a $(\boldsymbol{l} , \boldsymbol{r} )$-derivation if $\boldsymbol{d}_{p+q}=\boldsymbol{d}_{p} \otimes ' \boldsymbol{r} _{q} + \boldsymbol{l}_{p} \otimes ' \boldsymbol{d}_{q}$ for any $p,q \geq 0$ and $(\mathbf{1} ,\mathbf{1} )$-derivation is just a derivation, 
where $\mathbf{1}$ denotes the identity. 
A contracting homotopy $\boldsymbol{h}$ may be a $(\mathbf{1} , \boldsymbol{\iota \pi })$-derivation. 
We thus find their defining properties as follows 
\begin{align*}
\Delta ' \boldsymbol{f} & = 
\big{(} \boldsymbol{f} \otimes ' \boldsymbol{f} \big{)} \, \Delta ' \, , 
\\ 
\Delta ' \boldsymbol{d} & =
\big{(} \boldsymbol{d} \otimes ' \mathbf{1} + \mathbf{1} \otimes ' \boldsymbol{d} \big{)} \, \Delta '  \, , 
\\ 
\Delta ' \boldsymbol{h} & = 
\big{(} \boldsymbol{h} \otimes ' \boldsymbol{\iota \pi } + \mathbf{1} \otimes ' \boldsymbol{h} \big{)} \, \Delta '  \, . 
\end{align*} 
Actually, 
the property of contracting homotopy $\boldsymbol{h}$ is sufficient for our purpose. 
One could use a weaker condition, 
a pseudo-derivative condition of $\boldsymbol{h}$, 
which is given by 
\begin{align*}  
\big{(} \boldsymbol{h} \otimes ' \boldsymbol{h} \big{)} \Delta '  
= \big{(} \boldsymbol{h} \otimes ' \mathbf{1} - \mathbf{1} \otimes ' \boldsymbol{h}  \big{)} \Delta ' \boldsymbol{h} 
= - \big{(} \Delta ' \boldsymbol{h} \big{)} \big{(} \boldsymbol{h} \otimes ' \mathbf{1} - \mathbf{1} \otimes ' \boldsymbol{h}  \big{)}  
\, .
\end{align*}

\subsubsection*{Perturbation lemma for $A_{\infty }$}

Let $(\cH \overset{\boldsymbol{\pi }}{\underset{\boldsymbol{\iota }}{\scalebox{1.5}[1]{$\rightleftarrows $}}} \cL , \boldsymbol{h} )$ be a coalgebra contraction. 
When $\boldsymbol{\Delta }$ is a coderivation, 
$\boldsymbol{\iota _{\Delta }}$ and $\boldsymbol{\pi _{\Delta }}$ are morphisms of graded coalgebras and $\boldsymbol{Q_{\Delta }} = \boldsymbol{Q} + \boldsymbol{\Delta }$ and $\boldsymbol{q_{\Delta }}$ are coderivations. 
Then, 
we obtain a coalgebra version of the homological perturbation lemma by putting it together.\footnote{This is why a thick map is used. 
This useful notation will simplify the application of the lemma. } 
When a coderivation $\boldsymbol{\Delta }$ satisfies $(\boldsymbol{Q} + \boldsymbol{\Delta })^{2} =0$, 
the lemma is transferred to $A_{\infty }$. 

\vspace{2mm} 

We write $\boldsymbol{M} \equiv \{ \boldsymbol{Q} , \boldsymbol{M_{n}} \} _{n>1}$ for an $A_{\infty }$ structure of $\cT (\cH )$; 
we write $\boldsymbol{m} \equiv \{ \boldsymbol{q} , \boldsymbol{m_{n}} \} _{n>1}$ for an $A_{\infty }$ structure of $\cT (\cL )$. 
Let us consider a contraction of $A_{\infty }$ algebras 
\begin{subequations} 
\begin{align}
\label{before}
\boldsymbol{h } \, 
\rotatebox{-70}{$\circlearrowright $} \, 
\big{(} \cT (\cH ) , \boldsymbol{M} \big{)} 
\hspace{3mm} 
\overset{\boldsymbol{\pi } }{\underset{\boldsymbol{\iota }}{
\scalebox{2}[1]{$\rightleftarrows $}
}}
\hspace{3mm} 
\big{(} \cT ( \cL ) , \boldsymbol{m} \big{)} 
 \, . 
\end{align}
A coderivation $\boldsymbol{\Delta }$ is a perturbation for $\boldsymbol{M}$ when $\boldsymbol{M_{\Delta } } = \boldsymbol{M} + \boldsymbol{\Delta } $ is nilpotent, 
namely, 
$\boldsymbol{\Delta M} + \boldsymbol{M \Delta } + (\boldsymbol{\Delta })^{2} = 0$. 
We assume that a thick map $\mathbf{A}$ can be defined via the following \textit{recursive} relation
\begin{align*}
\mathbf{A} = \boldsymbol{\Delta } - \boldsymbol{\Delta h } \, \mathbf{A} \, . 
\end{align*} 
Then, 
because of the lemma, 
we obtain a contraction of $A_{\infty }$ algebras  
\begin{align}
\label{after}
\boldsymbol{h_{\Delta } } \, 
\rotatebox{-70}{$\circlearrowright $} \, 
\big{(} \cT (\cH ) , \boldsymbol{M} + \boldsymbol{\Delta } \big{)} 
\hspace{3mm} 
\overset{\boldsymbol{\pi _{\Delta }} }{\underset{\boldsymbol{\iota _{\Delta }}}{
\scalebox{3}[1]{$\rightleftarrows $}
}}
\hspace{3mm} 
\big{(} \cT ( \cL ) , \boldsymbol{m} + \boldsymbol{\pi \mathbf{A} \iota } \big{)} 
\, , 
\end{align}
\end{subequations} 
where the perturbed data are defined by the \textit{recursive} relations 
\begin{subequations} 
\begin{align}
\label{injection}
\boldsymbol{\iota _{\Delta } } & = \boldsymbol{\iota } - \boldsymbol{h \Delta \, \iota _{\Delta } } \, , 
\\ \label{projection}
\boldsymbol{\pi  _{\Delta }} & = \boldsymbol{\pi } - \boldsymbol{\pi _{\Delta } \Delta h } \, , 
\\ 
\boldsymbol{h_{\Delta }} & = \boldsymbol{h} - \boldsymbol{h_{\Delta } \Delta h } \, . 
\end{align}
\end{subequations} 
It is the homological perturbation lemma for $A_{\infty }$---a useful tool for describing the reduction of gauge symmetry in off-shell interacting theory, 
which we will see in section 4.

\subsection{Feynman graphs and minimal model}

As we will see, 
when the kinetic operator $\boldsymbol{Q}$ and the vertices $\boldsymbol{M}_{2} + \cdots $ of interacting field theory have a homotopy algebraic structure $\boldsymbol{M} = \boldsymbol{Q} + \boldsymbol{M}_{2} + \dots $, 
the homological perturbation describes the process of path-integrating-out the fields living in $(\boldsymbol{1} -\boldsymbol{ \iota \pi } )\, \cH $ of (\ref{before}). 
Because of the recursive definitions of $\boldsymbol{A}$ and the perturbed data (\ref{injection}-c), 
$\boldsymbol{A}$ is nothing but the Feynman graphs if $\boldsymbol{h}$ is a propagator and $\boldsymbol{\Delta }$ denotes the vertices. 
Thus, 
effective field theories can be obtained via the homological perturbation. 

\vspace{2mm}

The classical $S$-matrix is a typical example, 
which gives a minimal model of $A_{\infty }$.\footnote{When a given $A_{\infty }$ structure $\boldsymbol{M}$ has no linear part $\boldsymbol{M_{1}}$, 
it is called as minimal: 
$\bM _{\mathrm{min} }$ of (\ref{S-matrix}) is minimal. 
In string field theory, 
more relaxed notions, 
such as ``almost minimal'' in \cite{Konopka:2015tta}, 
may be useful. }  
We write $\cH _{\mathrm{phys} }$ for the space of the physical states of the \textit{free} theory, 
which is the cohomology of the BRST operator $Q$. 
Let us consider a contraction of $A_{\infty }$ algebras 
\begin{subequations} 
\begin{align}
\label{asymptotic}
\boldsymbol{h } \, 
\rotatebox{-70}{$\circlearrowright $} \, 
\big{(} \cT (\cH ) , \boldsymbol{Q} \big{)} 
\hspace{3mm} 
\overset{\boldsymbol{\pi } }{\underset{\boldsymbol{\iota }}{
\scalebox{2}[1]{$\rightleftarrows $}
}}
\hspace{3mm} 
\big{(} \cT ( \cH _{\mathrm{phys} } ) , \boldsymbol{0} \big{)} 
 \, , 
\end{align} 
and assume that $\boldsymbol{h}^{2}= \boldsymbol{0} $ and $\boldsymbol{h} \, \boldsymbol{\Pi } = \boldsymbol{\Pi } \, \boldsymbol{h} = \boldsymbol{0}$ where $\boldsymbol{\Pi } \equiv \boldsymbol{\iota }  \boldsymbol{\pi }$. 
One may regard this $\boldsymbol{\Pi }$ as a projector onto the on-shell states of the free theory or asymptotic string fields. 
In the Siegel gauge, 
$\boldsymbol{h}$ is given by the propagator $b_{0}L_{0}^{-1}$ having poles on $\boldsymbol{\Pi } \, \cH$, 
which may be cast as $h = b_{0} L_{0}^{-1} (1 - e^{-\infty L_{0}})$ and $\Pi  = e^{-\infty L_{0}}$. 
As a specific case of the homological perturbation (\ref{after}), 
the minimal model is obtained by taking interacting terms $\boldsymbol{\Delta _{\mathrm{min}} } \equiv \boldsymbol{M} - \boldsymbol{Q}$ as the perturbation to (\ref{asymptotic}). 
The perturbed differential $\boldsymbol{Q_{\Delta _{\mathrm{min}} }} \equiv \boldsymbol{M}$ is just the $A_{\infty }$ structure of the interacting field theory.  
%
The classical $S$-matrix is given by the right hand side of the perturbed $A_{\infty }$ data 
\begin{align}
\boldsymbol{h _{\Delta _{\mathrm{min} } } } \, 
\rotatebox{-70}{$\circlearrowright $} \, 
\big{(} \cT (\cH ) , \boldsymbol{M} \big{)} 
\hspace{3mm} 
\overset{\boldsymbol{\pi _{\Delta _{\mathrm{min} } }} }{\underset{\boldsymbol{\iota _{\Delta _{\mathrm{min} } } }}{
\scalebox{4.5}[1]{$\rightleftarrows $}
}}
\hspace{3mm} 
\big{(} \cT ( \cH _{\mathrm{phys} } ) , \bM _{\mathrm{min} } \big{)} 
 \, . 
\end{align} 
\end{subequations} 
It is called as a minimal model of $A_{\infty }$, 
which has no gauge degree because of $\cH _{\mathrm{phys}}$. 
Note that $\bM _{\mathrm{min} }$ itself is nilpotent and it may generate gauge symmetry if one consider some state space $\cL $ with relaxed conditions instead of $\cH _{\mathrm{phys}}$. 
%
The $A_{\infty }$ structure of the minimal model $\bM _{\mathrm{min} } \equiv \boldsymbol{0_{\Delta _{\mathrm{min} } } }$ takes the following form  
\begin{align}
\label{S-matrix}
\bM _{\mathrm{min} } 
= \boldsymbol{\pi }
\Big[ \boldsymbol{Q} + \frac{1}{1+ \boldsymbol{\Delta _{\mathrm{min} } h } } \boldsymbol{\Delta _{\mathrm{min} } } \Big] 
 \boldsymbol{\iota } 
 = \boldsymbol{\pi _{\Delta _{\mathrm{min}} } } \, \boldsymbol{M} \, \, \boldsymbol{\iota _{\Delta _{\mathrm{min}} } } \, 
\end{align}
where $\boldsymbol{\pi } \, \boldsymbol{Q} \, \boldsymbol{\iota } = \boldsymbol{0}$ and $\boldsymbol{\Delta _{\mathrm{min} } } (\mathbf{1} + \boldsymbol{h \Delta _{\mathrm{min} } }  )^{-1} = (\mathbf{1} + \boldsymbol{\Delta _{\mathrm{min} } h }  )^{-1}  \boldsymbol{\Delta _{\mathrm{min} } } $.
We find that $\bM _{\mathrm{min} }$ is a coderivation satisfying (\ref{cod Q}) 
on the tensor algebra of the cohomology $\cT (\cH _{\mathrm{phys} })$. 
We thus obtain the $(\boldsymbol{\Pi } , \boldsymbol{\Pi })$-derivation $\boldsymbol{\iota } \, \bM _{\mathrm{min} } \, \boldsymbol{\pi }$ acting on $\cT ( \cH )$ as follows 
\begin{align}
\Delta ' \, \boldsymbol{\iota } \, \bM _{\mathrm{min} } \, \boldsymbol{\pi } 
= \big{(} \, \boldsymbol{\iota } \, \bM _{\mathrm{min} } \, \boldsymbol{\pi} \, \otimes ' \boldsymbol{\Pi } 
+ \boldsymbol{\Pi } \otimes ' \, \boldsymbol{\iota } \, \bM _{\mathrm{min} } \, \boldsymbol{\pi } \, \big{)} \Delta ' \, . 
\end{align}
It defines multi-linear maps acting on the physical states of the free theory, 
or on-shell asymptotic string fields, 
and is graphically same as the Feynman graphs reproducing amplitudes. 
(See also \cite{Kajiura:2003ax, Konopka:2015tta, Erler:2016rxg, Doubek:2017naz} or references of \cite{Doubek:2017naz}.) 
The perturbed injection $\boldsymbol{\iota _{\Delta _{\mathrm{min}} } }$, 
the perturbed projection $\boldsymbol{\pi _{\Delta _{\mathrm{min}}}  }$ 
and the perturbed contracting homotopy $\boldsymbol{h_{\Delta _{\mathrm{min}}} }$ are given by 
\begin{align*}
\boldsymbol{\iota _{\Delta _{\mathrm{min}} } } = \frac{1}{1 + \boldsymbol{h \Delta _{\mathrm{min}} } } \boldsymbol{\iota } \, , 
\hspace{5mm} 
\boldsymbol{\pi _{\Delta _{\mathrm{min}}}  }  = \boldsymbol{\pi } \frac{1}{1 + \boldsymbol{ \Delta _{\mathrm{min}} h} } \, , 
\hspace{5mm}  
\boldsymbol{h_{\Delta _{\mathrm{min} } } } = \boldsymbol{h} \frac{1}{1 + \boldsymbol{ \Delta _{\mathrm{min} } h } }  \, .
\end{align*} 
These $\boldsymbol{\iota _{\Delta _{\mathrm{min}}}  }$ and $\boldsymbol{\pi _{\Delta _{\mathrm{min}} } }$, 
as well as $\iota $ and $\pi $, 
are cohomomorphisms\footnote{It may induce a nonlinear field redefinition between the original and asymptotic string fields. } since they satisfy (\ref{coh iota}) and (\ref{coh pi}). 
%
%
%
Likewise, 
these $\boldsymbol{\iota _{\Delta _{\mathrm{min}}}  }$ and $\boldsymbol{\pi _{\Delta _{\mathrm{min}} } }$ satisfy the projection properties 
\begin{align*}
(\boldsymbol{ \iota _{\Delta _{\mathrm{min} }} \pi _{\Delta _{\mathrm{min} }} } )^{2}= 
\boldsymbol{ \iota _{\Delta _{\mathrm{min} }} \pi _{\Delta _{\mathrm{min} }} } \, , 
\hspace{5mm} 
(\boldsymbol{ \pi _{\Delta _{\mathrm{min} }} \iota _{\Delta _{\mathrm{min} }} } )^{2}= 
\boldsymbol{ \pi _{\Delta _{\mathrm{min} }} \iota _{\Delta _{\mathrm{min} }} } \, . 
\end{align*}
Since $\boldsymbol{h_{\Delta _{\mathrm{min} } } } $ satisfies (\ref{coc h}), 
we find the Hodge type decomposition on $\cT (\cH )$ 
\begin{align*}
\mathbf{1} - \boldsymbol{\iota _{\Delta _{\mathrm{min} }} \pi _{\Delta _{\mathrm{min} }} } = 
\boldsymbol{M} \, \boldsymbol{h_{\Delta _{\mathrm{min} } } } + 
\boldsymbol{h_{\Delta _{\mathrm{min} } } } \, \boldsymbol{M} \, . 
\end{align*}
It defines a nonlinear decomposition of the unit: 
$\bM$-exact, 
$\boldsymbol{h_{\Delta _{\mathrm{min} } } }$-exact and on-shell states.

\section{Light-cone reduction}

Let $\Phi _{\mathrm{old}}$ be a light-cone string field of the light-cone formulation \cite{Kaku:1974zz}. 
The kinetic term of the light-cone formulation is given by 
\begin{align*}
S_{\mathrm{old} ,\, 2} [\Phi _{\mathrm{old} }] = \frac{1}{2} 
\int \cD [x^{+} , p^{+} ] \, \frac{p^{+} }{\pi } \, 
\La \Phi _{\mathrm{old} }( x^{+} , p^{+} ) , \, \frac{K^{lc} }{2 p^{+}} \, \Phi _{\mathrm{old} } (x^{+} , p^{+} ) \Ra _{X^{I} } \, . 
\end{align*}
We write $\cH _{\mathrm{old}} $ for the state space of the light-cone formulation: $\Phi _{\mathrm{old} } \in \cH _{\mathrm{old}}$. 
As we explained in the previous section, 
by applying the homological perturbation lemma with $d = \sum d_{n}$, 
the kinetic term of the Witten theory reduces to 
\begin{align*}
S_{2} [ \Psi _{\mathrm{lc} } ] 
= \frac{1}{2} \la \, \Psi _{\mathrm{lc} } \, , \,  c_{0} \, K^{lc} \, \Psi _{\mathrm{lc} } \, \ra  \, . 
\end{align*}
The string field $\Psi _{\mathrm{lc}} \in \cH _{\mathrm{lc} }$ and the kinetic operator $K^{lc}$ consists of physical excitations $\{ x^{\mu } , p_{\mu } , a^{I}_{n} \} _{n} $ for $0< I < 25$.  
The kinetic term has no $a^{\pm }_{n}$-excitation and no non-zero ghost excitation from the Fock vacuum. 
We thus find $K^{lc} \, \cH _{\mathrm{lc} } \subset \cH _{\mathrm{lc} }$. 
Since the Fock vacuum takes $| \Omega \rangle \equiv c_{1} | 0 \rangle $ and the $bc$-ghost number anomaly implies $ \langle \Omega | c_{0} | \Omega \rangle \equiv \langle 0 | c_{-1} c_{0} c_{1} | 0 \rangle \not= 0$, 
we can map from $\Psi _{\mathrm{lc}} \in \cH _{\mathrm{lc} }$ to $\Phi _{\mathrm{old}} \in \cH _{\mathrm{old}}$ by $\Psi _{\mathrm{lc} } = c_{1} \Phi _{\mathrm{old} } (x^{+} , p^{+}) $. 
Hence, 
for any states $A , B \in \cH _{\mathrm{lc} }$, 
we find the equivalence of the inner products 
 \begin{align*}
\la \,  A , \, c_{0} B \, \ra = 
\frac{1}{2 \pi } \int \cD [x^{+} , p^{+} ] \, 
\La A ( x^{+} , p^{+} ) , \, B (x^{+} , p^{+} ) \Ra _{X^{I} } 
\, . 
\end{align*}

\vspace{2mm}

Now, we have the Hodge type decomposition of the state space $\cH _{\mathrm{cov} }$ as follows
\begin{align}
\label{Hodge decomposition}
\mathbf{1} - \boldsymbol{\Pi } = \boldsymbol{d} \, \boldsymbol{h} + \boldsymbol{h} \, \boldsymbol{d} \, , 
\end{align}
where $\boldsymbol{h}$ is given by (\ref{lc h}) and $\boldsymbol{\Pi }$ denotes a projector onto the state space $\cH _{\mathrm{lc}} = \Pi \, \cH _{\mathrm{cov} }$ of the light-cone reduction.  
It also works well on the Fock space $\cT (\cH _{\mathrm{cov} })$ with (\ref{thick}). 
Using it, 
we elucidate how the light-cone string field appears in the interacting theory.

\subsection{Light-cone reduction and $A_{\infty }$ structure} 

Witten's open string field theory has a cubic action 
\begin{align}
\label{Witten}
S [\Psi ] = \frac{1}{2} \la \Psi , \, Q \, \Psi \ra + \frac{1}{3} \la \Psi , \, m_{2} \big{(} \Psi , \Psi \big{)} \ra \, . 
\end{align}
Let us consider the following redefinitions of the string field $\Psi $ and the star product $m_{2}$, 
\begin{align}
\label{redefinitions}
\Psi _{\mathrm{cov}} = \cU \, \Psi  \, , \hspace{5mm} 
\boldsymbol{m^{\mathrm{cov}}_{\, 2}} = \boldsymbol{\cU } \, \boldsymbol{m_{2}} \, \boldsymbol{\cU }^{-1}  \, . 
\end{align}
Namely, 
we set $m^{\mathrm{cov} }_{\, 2} = {\cU } \, m_{2} ( \cU ^{-1} \otimes \cU ^{-1} ) $ as a bilinear map. 
Hence, 
the cubic vertex $m^{\mathrm{cov}}_{2}$ remains associative and is exactly equivalent to the original star product $m_{2}$. 
We find 
\begin{align}
\label{LC Witten}
S [\Psi _{\mathrm{cov}} ] 
= \frac{1}{2} \la \, \Psi _{\mathrm{cov}} , \, ( c_{0} \, K^{lc} + d ) \, \Psi _{\mathrm{cov}} \, \ra 
+ \frac{1}{3} \la \, \Psi _{\mathrm{cov}} , \, m^{\mathrm{cov}}_{\, 2} \big{(} \Psi _{\mathrm{cov}} , \Psi _{\mathrm{cov}} \big{)} \, \ra \, . 
\end{align}
The string field $\Psi _{\mathrm{cov}} \in \cH _{\mathrm{cov} }$ has ghost number $1$ and includes gauge degrees. 
The action $S[\Psi _{\mathrm{cov} } ]$ is invariant under the gauge transformation 
\begin{align*}
\delta \Psi _{\mathrm{cov}} = ( c_{0} \, K^{lc} + d \, ) \, \Lambda _{\mathrm{cov} } 
+ m^{\mathrm{cov} }_{\, 2} \big{(} \Psi _{\mathrm{cov}} , \Lambda _{\mathrm{cov}} \big{)}
+ m^{\mathrm{cov} }_{\, 2} \big{(} \Lambda _{\mathrm{cov}} , \Psi _{\mathrm{cov}} \big{)} \, , 
\end{align*}
where $\Lambda _{\mathrm{cov}} \in \cH _{\mathrm{cov} }$ is a gauge parameter field carrying ghost number $0$. 
By defining $m^{\mathrm{cov} }_{\, 1} \equiv c_{0} \, K^{lc} + d$, 
we find that as the Witten theory, 
the following coderivation becomes nilpotent and gives a cyclic $A_{\infty }$ structure 
\begin{align}
\label{cov A-infinity}
\boldsymbol{m^{\mathrm{cov} }} \equiv \boldsymbol{m^{\mathrm{cov} }_{\, 1}} + \boldsymbol{m^{\mathrm{cov} }_{\, 2}} \, . 
\end{align}
The original $A_{\infty }$ relations of (\ref{Witten}) provide the nilpotency $(\boldsymbol{Q})^{2}=0$, 
the Leibniz rule $\ld \boldsymbol{Q} , \boldsymbol{m_{2}} \rd = 0$ and the associativity of the star product $(\boldsymbol{m_{2} })^{2}= 0$. 
In terms of (\ref{cov A-infinity}), 
these relations correspond to 
\begin{subequations} 
\begin{align}
( \boldsymbol{c_{0} K^{lc}} )^{2} = ( \boldsymbol{d} )^{2} = \Ld \boldsymbol{d} , \boldsymbol{c_{0} K^{lc}} \Rd = 0 \, , 
\\ 
\Ld \boldsymbol{d} , \boldsymbol{m^{\mathrm{cov}}_{\, 2} } \Rd 
+ \Ld \boldsymbol{c_{0} K^{lc}} , \boldsymbol{m^{\mathrm{cov}}_{\, 2} } \Rd = 0 \, , 
\\ 
( \boldsymbol{m^{\mathrm{cov} }_{\, 2}} )^{2} = 0 \, , 
\end{align}
\end{subequations} 
respectively. 
Note that although the physical and unphysical excitations are completely split in the above expression, 
it remains covariant theory yet. 

\vspace{2mm} 

We can expand the string field as $\Psi _{\mathrm{cov}} = \psi + c_{0} \chi $ and 
the action (\ref{LC Witten}) becomes 
\begin{align*}
S [\psi + c_{0} \chi ] & = \frac{1}{2} \la \psi , \,  c_{0} K^{lc} \, \psi  \ra 
+ \frac{1}{3} \la \psi  , \, m^{\mathrm{cov} }_{\, 2} \big{(} \psi  , \psi  \big{)} \ra  
+ \la c_{0} \chi , \, m^{\mathrm{cov} }_{\, 2} \big{(} \psi  , \psi  \big{)} \ra 
\no & \hspace{5mm} 
+ \la c_{0} \chi , \, d \, \psi \ra 
+ \la \psi  , \, m^{\mathrm{cov} }_{\, 2} \big{(} c_{0} \chi  , c_{0} \chi  \big{)} \ra 
+ \frac{1}{3} \la c_{0} \chi  , \, m^{\mathrm{cov} }_{\, 2} \big{(} c_{0} \chi  , c_{0} \chi  \big{)} \ra \, . 
\end{align*}
Roughly speaking, 
the covariant theory reduces to the light-cone theory by integrating out $\chi$ and by solving its equations of motion 
\begin{align*}
c_{0} \Big[ d \, \psi + m^{\mathrm{cov} }_{\, 2} ( \psi , c_{0} \chi ) + m^{\mathrm{cov} }_{\, 2} (c_{0} \chi , \psi ) 
+ m^{\mathrm{cov} }_{\, 2} (\psi , \psi ) + m^{\mathrm{cov} }_{\, 2} ( c_{0} \chi , c_{0} \chi ) \Big] = 0 \, . 
\end{align*}
This type of reduction can be performed by the homological perturbation lemma.\footnote{The path-integral-based understanding of homological perturbation lemma was investigated in the early days. 
See textbooks such as \cite{Henneaux}. 
For recent works, 
references in \cite{Doubek:2017naz} may be helpful. } 
It is an exact procedure to reduce nonlinear gauge symmetry and provides a closed form expression of the resultant theory.
We thus consider the lemma for $A_{\infty }$, 
or the minimal model.

\vspace{2mm} 

Let us explain the light-cone reduction of Witten's string field theory. 
In the Witten theory, 
because of (\ref{Hodge decomposition}) and (\ref{redefinitions}), 
any covariant string field $\Psi _{\mathrm{cov} } \in \cH _{\mathrm{cov} }$ can be decomposed as $\Psi _{\mathrm{cov} } = d \, h \, \Psi _{\mathrm{cov} } + h \, d \, \Psi _{\mathrm{cov} } + \Pi \Psi _{\mathrm{cov} }$ with $\Pi \, \Psi _{\mathrm{cov} } = \Psi _{\mathrm{lc} } \in \cH _{\mathrm{lc}}$. 
We showed that for the free theory of covariant string fields, 
the light-cone reduction is obtained by the contraction of $A_{\infty }$ algebras 
\begin{subequations} 
\begin{align}
\boldsymbol{h } \, 
\rotatebox{-70}{$\circlearrowright $} \, 
\big{(} \cT (\cH _{\mathrm{cov} }) , \boldsymbol{m^{\mathrm{cov} }_{\, 1} } 
\big{)} 
\hspace{3mm} 
\overset{\boldsymbol{\pi } }{\underset{\boldsymbol{\iota }}{
\scalebox{2}[1]{$\rightleftarrows $}
}}
\hspace{3mm} 
\big{(} \cT ( \cH _{\mathrm{lc} } ) , \boldsymbol{c_{0}  K^{lc} }  \big{)} 
 \, , 
\end{align}
which is an alternative proof of the no-ghost theorem of covariant strings.\footnote{
It is nothing but the result of the perturbation $\Delta = c_{0}K^{lc}$ for $h \, 
\rotatebox{-70}{$\circlearrowright $} \, 
\big( \cT (\cH _{\mathrm{cov} } ) , d \big) 
\, 
\rightleftarrows 
\, 
\big( \cT (\cH _{\mathrm{lc}} ) , 0 \big{)}$ 
with (\ref{Hodge decomposition}). 
}  
In order to include interactions, 
we consider the perturbation $\boldsymbol{\Delta } \equiv \boldsymbol{m^{\mathrm{cov} }_{2} } $ for the above contraction of $A_{\infty }$. 
The perturbed $A_{\infty }$ data are 
\begin{align}
\boldsymbol{h _{\Delta } } \, 
\rotatebox{-70}{$\circlearrowright $} \, 
\big{(} \cT (\cH _{\mathrm{cov} }) , \boldsymbol{m^{\mathrm{cov} } } \big{)} 
\hspace{3mm} 
\overset{\boldsymbol{\pi _{\Delta }} }{\underset{\boldsymbol{\iota _{\Delta } }}{
\scalebox{3}[1]{$\rightleftarrows $}
}}
\hspace{3mm} 
\big{(} \cT ( \cH _{\mathrm{lc} } ) , \boldsymbol{m^{lc} } \big{)} 
 \, , 
\end{align} 
\end{subequations} 
where the $A_{\infty }$ structure of the reduced theory is given by 
\begin{align}
\label{def of lc}
\boldsymbol{m^{lc} }  
\equiv \boldsymbol{\pi }
\Big[ \boldsymbol{c_{0} K^{lc} } + \frac{1}{1+ \boldsymbol{ m^{\mathrm{cov} }_{\, 2} \, h } } \boldsymbol{m^{\mathrm{cov}}_{\, 2} } \Big] 
 \boldsymbol{\iota } \, . 
\end{align} 
The cyclicity is manifest because of $\boldsymbol{h}$. 
The contraction preserves the cohomology and two $A_{\infty }$ structures $\boldsymbol{m^{\mathrm{cov} }}$ and $\boldsymbol{m^{lc}}$ give the same physical spectrum. 
Hence, 
because of the lemma, 
the reduced theory reproduces the same string amplitudes as Witten's string field theory.\footnote{The minimal model theorem for $A_{\infty }$ ensures uniqueness of the minimal model of these cyclic $A_{\infty }$ algebras. 
In other words, 
(\ref{LC Witten}) and (\ref{lc red action}) have the same S-matrix at the tree level. 
} 
Note that $\boldsymbol{m^{lc}} : \cT ( \cH _{\mathrm{lc} } ) \rightarrow \cT (\cH _{\mathrm{lc} })$ and $\boldsymbol{m^{lc} }$ defines the vertices of the light-cone string field theory via 
\begin{align*}
\boldsymbol{m^{lc}} = \boldsymbol{m^{lc}_{1}} + \boldsymbol{m^{lc}_{2}} + \boldsymbol{m^{lc}_{3}} + \cdots + \boldsymbol{m^{lc}_{n} } + \cdots \, . 
\end{align*} 
One can identify $\boldsymbol{m^{lc}}$ acting on $\cT (\cH _{\mathrm{lc} })$ with $\boldsymbol{\iota } \, \boldsymbol{m^{lc}} \, \boldsymbol{\pi }$ acting on the physical subspace of $\cT (\cH _{\mathrm{cov} } )$, 
for which we also write $\cT (\cH _{\mathrm{lc} } ) \subset \cT (\cH _{\mathrm{cov} } )$. 
By expanding (\ref{def of lc}), 
we find the explicit forms of these $A_{\infty }$ multilinear maps $\{ m^{lc}_{n} \}_{n=1}^{\infty }$ acting on $\cT (\cH _{\mathrm{cov} })$ as follows 
\begin{subequations} 
\begin{align}
\label{lc 1-product}
m^{lc}_{1} & \equiv c_{0} \, K^{lc} \, , 
\\ \label{lc 2-product}
m^{lc}_{2} & \equiv \Pi \, m^{\mathrm{cov} }_{\, 2} \big{(} \Pi  \otimes \Pi  \big{)} \, , 
\\ \label{lc 3-product}
m^{lc}_{3} & \equiv - \Pi \, \Big[ m^{\mathrm{cov} }_{\, 2} ( h \, m^{\mathrm{cov} }_{\, 2} \otimes 1) 
+ m^{\mathrm{cov} }_{\, 2} ( 1 \otimes h \, m^{\mathrm{cov} }_{\, 2} ) \Big] \Pi ^{\otimes 3} \, , 
\\ 
& \hspace{1.5mm} \vdots 
\no \label{lc n-product}
m^{lc}_{n} & \equiv(-)^{n} \Pi \, \Big[ \boldsymbol{m^{\mathrm{cov} }_{\, 2} } \frac{1}{1 - \boldsymbol{h \, m^{\mathrm{cov} }_{\, 2} }} \Big] \, \Pi ^{\otimes n}  \, , 
\\ \nonumber 
& \hspace{1.5mm} \vdots 
\end{align}
\end{subequations} 
where $\Pi ^{\otimes n} $ maps $(\cH _{\mathrm{cov} } )^{\otimes m}$ to $(\cH _{\mathrm{lc} } )^{\otimes n}$ iff $m=n$, otherwise to $0$. 
Note that these are tree graphs and equal to the classical parts of the vertices of an effective action for the Witten theory if one regards $h$ as a propagator, 
in which gauge degrees are integrated out instead of high-energy physical degrees as (\ref{effective}). 
We obtain the reduced action 
\begin{align}
\label{lc red action}
S_{lc} [\Psi _{\mathrm{lc} } ] = \frac{1}{2} \la \Psi _{\mathrm{lc} } , \,  c_{0} K^{lc} \, \Psi _{\mathrm{lc} }  \ra 
+ \sum_{n>1} \frac{1}{n+1} \la \Psi _{\mathrm{lc} } , \, m^{lc}_{n} \big{(} \Psi _{\mathrm{lc} } , \dots , \Psi _{\mathrm{lc} } \big{)} \ra  \, , 
\end{align}
whose kinetic term is just equivalent to that of the old light-cone formulation. 
Although the resultant theory is consistent as a light-cone theory, 
it necessitates an infinite number of vertices (\ref{lc 1-product}-d) unlike the old light-cone formulation \cite{Kaku:1974zz}. 
The action (\ref{lc red action}) takes an $A_{\infty }$ form and it satisfies the $A_{\infty }$ relations $(\boldsymbol{m^{lc}} )^{2} = 0$ because of the homological perturbation. 
Note however that the light-cone reduction kills all unphysical states. 
The string field $\Psi _{\mathrm{lc}}$ has no gauge degree, 
\begin{align*}
\delta \Psi _{\mathrm{lc} } = 0 \, . 
\end{align*}

\subsection{Reduction of gauge symmetry and covariance}

Although our light-cone theory (\ref{lc red action}) has no gauge degree, 
there exists a potential $A_{\infty }$ structure $(\boldsymbol{m^{lc} })^{2} = 0$ as the old light-cone formulation. 
Namely, 
for each $n >0$, 
we have 
\begin{align}
\sum _{k+l =n} m^{lc}_{l+1} \big{(}  \dots \, m^{lc}_{k} \, \dots \big{)} = 0 \, . 
\end{align}
If there was\footnote{Such a ghost number $0$ state must include some BRST quartet excitation, 
which is projected out in the process of the reduction. 
In other words, 
the theory is already gauge-fixed. } any ghost number $0$ state $\lambda $ in $\cH _{\mathrm{lc} }$, 
it could generate the gauge transformation 
\begin{align}
\delta \Psi _{\mathrm{lc} } = c_{0} K^{lc} \lambda + m^{lc}_{2} (\Psi _{\mathrm{lc} } , \lambda ) + m^{lc}_{2} ( \lambda , \Psi _{\mathrm{lc} } ) + \cdots \, . 
\end{align} 
One can restore this type of gauge symmetry by adding trivial BRST quartets, 
which we removed in the light-cone reduction. 
As we explain, 
it recovers the covariance. 

\vspace{2mm} 

Let us consider state spaces spanned by trivial BRST quartets $\cG _{n} \equiv \big{\{ } a^{\pm }_{\pm k} , b_{\pm k} , c_{\pm k}  \big{\} }_{k=1}^{n}$. 
There is a sequence of state spaces intermediating the light-cone and covariant theories: 
\begin{align*}
\cH _{\mathrm{lc}} 
\hspace{1mm} \subset \hspace{1mm} 
\cH _{\mathrm{lc}} \otimes \cG _{1} 
\hspace{1mm} \subset \hspace{1mm} \cdots \hspace{1mm} 
\subset \cH _{\mathrm{lc}} \otimes \cG _{n} 
\hspace{1mm} \subset \hspace{1mm} \cdots \hspace{1mm} 
\subset \cH _{\mathrm{cov}} \equiv \cH _{\mathrm{lc}} \otimes \cG _{\infty } \, . 
\end{align*}
The covariance is restored by adding the $\cG _{n}$-quartet's excitations successively. 
We write $\Pi _{(n)}$ for a projector onto $\cH _{\mathrm{lc} } \otimes \cG _{n}$, namely, $\Pi _{(n)} \cH _{\mathrm{cov} } \equiv \cH _{\mathrm{lc} } \otimes \cG _{n}$\,. 
We find the Hodge type decomposition $\mathbf{1} - \boldsymbol{\Pi _{(n)}} = \boldsymbol{d_{(n)}} \, \boldsymbol{h_{(n)} } + \boldsymbol{h_{(n)}} \, \boldsymbol{d_{(n)} }$ by using (\ref{d and h}) and 
\begin{align*}
d_{(n)} \equiv d - \sum_{k =1}^{n} d_{k} - \sum_{k=1}^{n} d_{-k} \, ,  
\hspace{5mm} 
h_{(n)} \equiv h - \sum_{k=1}^{n} h_{k} - \sum_{k=1}^{k} h_{-k} \, . 
\end{align*} 
These subtracting operators $\{ d_{k} , d_{-k} \}_{k=1}^{n}$ generate gauge symmetry on $\Pi _{(n)} \cH _{\mathrm{cov} }$ and the $\cG _{n}$-quartet excitations always appear in its gauge parameter field $\Lambda _{n} \in \Pi _{(n)} \cH _{\mathrm{cov} }$. 
The intermediate theories are covariant for some space-time fields and are light-cone type for the other space-time fields, 
in which covariant space-time fields have gauge degrees. 
A string field $\Psi _{n} \in \Pi _{(n)} \cH _{\mathrm{cov} }$ therefore has gauge degrees $\delta \Psi _{n} \in \Pi _{(n)} \cH _{\mathrm{cov} }$. 
We find the reduced (cyclic) $A_{\infty }$ structure  
\begin{align*}
\boldsymbol{m^{(n)}} \equiv \boldsymbol{\Pi _{(n)} } \Big[ \boldsymbol{c_{0} K^{lc} } 
+ \sum_{k = 1}^{n} \boldsymbol{d_{k}} 
+ \sum_{k=1}^{n} \boldsymbol{d_{-k} }  
+ \frac{1}{1 + \boldsymbol{m^{\mathrm{cov} }_{\, 2} } \boldsymbol{h_{(n)} } } \boldsymbol{m^\mathrm{cov}_{\, 2} }  \Big] \boldsymbol{\Pi _{(n) }} \, . 
\end{align*}
It defines a kinetic operator $m^{(n)}_{1} \equiv c_{0} K^{lc} + \sum _{k=1}^{n} ( d_{k} + d_{-k} ) $ and an $A_{\infty }$ string field theory\,. 
We obtain a \textit{gauge invariant} action $I_{n} [\Psi _{n} ]$ for the $n$-th intermediate theory 
\begin{align*}
I_{n} [\Psi _{n} ] 
= \frac{1}{2}
\La \Psi _{n} , \,  
\big{(} c_{0} K^{lc} 
+ \sum_{k=1}^{n} d_{k} 
+ \sum_{k=1}^{n} d_{-k}
\big{)} \, \Psi _{n}  \Ra 
+ \sum_{k=2}^{\infty } \frac{1}{k+1} 
\La \Psi _{n} , \, 
m^{(n)}_{k} \big{(} \Psi _{n} , \, ... \, , \Psi _{n} \big{)} \Ra  \, . 
\end{align*}
It is invariant under the gauge transformation 
\begin{align*}
\delta \Psi _{n} = \big{(} c_{0} K^{lc} + \sum_{k=1}^{n} d_{k} + \sum_{k=1}^{n} d_{-k} \big{)} \Lambda _{n} 
+ \sum_{k=1}^{\infty }  \sum_{\mathrm{cyclic}} m^{(n)}_{k+1} \big{(} \underbrace{\Psi _{n} , \, ...\, , \Psi _{n} }_{k}, \Lambda _{n}  \big{)} \, .  
\end{align*} 
One cannot get $\Lambda _{n}$ without quartet excitations: 
$\Lambda _{n} \notin \cH _{\mathrm{lc} }$ but $\Lambda _{n} \in \cH _{\mathrm{lc} } \otimes \cG _{n}$. 

\vspace{2mm} 

The theory $I_{1} [\Psi _{1} ]$ may give an interesting example, 
in which a string field $\Psi _{1} \in \cH _{\mathrm{lc}} \otimes \cG _{1}$ has the $\cG _{1}$-quartet excitations. 
The string field $\Psi _{1}$ consists of the tachyon $t (x)$, 
\textit{covariant} photon or Yang-Mills field $A_{\mu }(x)$ and transverse components of the massive higher-spin fields $\{ \varphi _{\mu \nu } (x) , \varphi _{\mu \nu \rho } , \cdots \}$. 
Now, 
a ghost number $0$ state $\Lambda _{1} \in \cH _{\mathrm{lc}} \otimes \cG _{1}$ exists because the state space includes the $\cG _{1}$-quartet excitations. 
It is a \textit{gauge invariant} theory and covariant just for Yang-Mills fields $A_{\mu } (x)$. 
Note that $A_{\mu } (x)$'s gauge degrees generate the $A_{\infty }$ type gauge transformation of the string field 
\begin{align*}
\delta \Psi _{1} = \big{(} c_{0} K^{lc} + d_{1} + d_{-1} \big{)} \Lambda _{1} 
+\sum_{n=1}^{\infty }  \sum_{\mathrm{cyclic}}  m^{(1)}_{n+1} \big{(} \underbrace{\Psi _{1} , \, ... \, , \Psi _{1} }_{n}, \Lambda _{1}  \big{)} \, .  
\end{align*}

\section{Conclusion}

In this paper, 
we proposed the light-cone reduction of covariant string field theory on the basis of the homological perturbation lemma for $A_{\infty }$. 
The resultant theory is consistent as a light-cone theory and its action takes an $A_{\infty }$ form (\ref{lc red action}), 
which seems to be different\footnote{In this paper, 
we considered a linear field-redefinition (\ref{redef}), 
which will correspond to a usual gauge-fixing condition of the perturbative field theory. 
Our results may imply that we necessitate a nonlinear field-redefinition or non-perturbative gauge-fixing in order to fill this gap. } from the old light-cone formulation \cite{Kaku:1974zz}. 
It would be noteworthy, 
for it implies that the old light-cone formulation contains some additional structure simplifying string field theory, 
which is missing in the covariant formulation based on the minimal world-sheet variables. 
\\
We also showed that the homological perturbation lemma indeed provides not only an exact gauge-fixing procedure taking into account interactions as (\ref{effective}), 
but also direct correspondence between different theories as (\ref{HPL}). 
It gives alternative treatment of the no-ghost theorem \cite{Kato:1982im} or similarity transformations of the BRST operator \cite{Aisaka:2004ga}. 
\\
We conclude this section with some comments on related topics.

\vspace{2mm} 

When the total central charge is nonzero, 
amplitudes depend on a world-sheet metric, 
which is taken to be flat along a propagator in the Witten theory. 
Although our theory is obtained from the Witten theory within the covariant formulation,  
its world-sheet metric becomes \textit{flat along light-cone diagrams} unlike the Witten theory.\footnote{The author would like to thank N.~Ishibashi for the advise which was new information for him. } 
It would be interesting to clarify this mechanism and compare our light-cone reduction with the earlier covariantized light-cone approach  \cite{Kugo:1987rq ,Siegel:1987ku}. 
Another interesting future direction is applying the lemma to the covariantized light-cone closed string field theory. 
These are in progress.

\vspace{2mm} 

The homological perturbation gives an alternative approach to (partial) gauge-fixing. 
It would be interesting to clarify how to apply the lemma to WZW-like string field theory, 
in which a pair or triplet of $A_{\infty } / L_{\infty }$ generates gauge degrees \cite{Matsunaga:2016, Erler:2017onq}. 

\vspace{2mm} 

One can understand Sen's Wilsonian effective action \cite{Sen:2016qap} in the same manner, 
in which the homological perturbation lemma should be applied to quantum $L_{\infty }$ or quantum BV. 
As shown by \cite{Doubek:2017naz}, 
a minimal model for a given quantum $L_{\infty }$ algebra can be constructed by applying the lemma, 
which naturally gives a Feynman diagram expansion and defines an effective theory. 
It enables us to construct an effective action satisfying the quantum BV master equation from a given gauge theory satisfying that.

\subsection*{Acknowledgments}

The author would like to thank Ted Erler for helpful and convenient discussions: 
this work was initiated in joint works with him and the main ideas of applying the minimal model are credited to Ted Erler. 
The author also thanks Nobuyuki Ishibashi, 
Yoichi Kazama, 
Hiroshi Kunitomo, 
Toru Masuda, 
Yuji Okawa and Martin Schnabl. 
This research has been supported in part by 
the Czech Science Foundation (GA$\check{\mathrm{C}}$R) grant 17-22899S.

\appendix

\section{On the homological perturbation lemma} 

We explain what the homological perturbation lemma (\ref{HPL}) is and why it is useful in gauge field theory. 
Appendix A.1 is devoted to a short proof of the lemma and appendix A.2 includes a few examples. 
As application to strings, 
we construct similarity transformations of the BRST operator by using the lemma in appendix A.3. 

\vspace{2mm} 

Let us consider a complex $(\cH , Q)$, a pair of a state space $\cH $ and a differential $Q$ acting on $\cH$. 
We write $\Pi $ for a restriction onto a subspace $\cL \subset \cH $ and $q = Q|_{\cL }$ for the differential acting on $\cL = \Pi \, \cH $. 
When two pairs $(\cH , Q)$ and $(\cL , q )$ have the same cohomology, 
we obtain Hodge type decomposition of the unit $1$ of $\cH $, 
\begin{align*}
1 = Q \, h + \, h \, Q + \Pi \, , 
\end{align*}
using a contracting homotopy operator $h$ for $Q$. 
It is helpful to distinguish $\cL $ from $\cH $ by separating $\Pi = \iota \, \pi $ into a natural projection $\pi $ and a natural injection $\iota $. 
Namely, 
we obtain $\pi \, \cH = \cL $, $\pi \, Q = q \, \pi $, $\iota \, \cL = \Pi \, \cH $ and $\iota \, q = Q \, \iota $ from the Hodge type decomposition, 
which gives a standard situation (\ref{SS}). 
It also implies the homotopy equivalence between two complexes, 
which is important for the lemma. 
A typical example is the Siegel gauge, 
in which $Q$ is the BRST operator, 
$h$ is a propagator $\frac{b_{0}}{L_{0}}$ and $\Pi $ is a projector onto the space of the physical states $\mathrm{Ker} [L_{0} ]$. 
The light-cone decomposition gives another example.

\vspace{2mm} 

The homological perturbation lemma provides us a correspondence of standard situations concretely, 
for which we write $\cU (t)$ in (\ref{HPL}) as follows. 
\begin{align}
\label{HPL}
h \, 
\rotatebox{-70}{$\circlearrowright $} \, 
\big{(} \cH , Q \big{)} 
\hspace{3mm} 
& \hspace{3mm} 
\overset{\pi }{\underset{\iota }{
\scalebox{4}[1]{$\rightleftarrows $}
}}
\hspace{3mm} 
\hspace{3mm} 
\big{(} \cL , q \big{)} 
\no 
{}^{\cU (t)} \hspace{1.5mm} 
\scalebox{1}[3]{$\uparrow $} \hspace{0mm} 
\scalebox{1}[3]{$\downarrow $} \hspace{2mm} 
{}^{\cU ^{-1}(t)}  
\hspace{-4mm} 
& 
\\ \nonumber 
h_{\Delta (t)} \, 
\rotatebox{-70}{$\circlearrowright $} \, 
\big{(} \cH , Q_{\Delta (t)} \big{)} 
&\hspace{3mm} 
\overset{\pi _{\Delta (t)}}{\underset{\iota _{\Delta (t)}}{
\scalebox{4}[1]{$\rightleftarrows $}
}}
\hspace{3mm} 
\big{(} \cL , q_{\Delta (t)} \big{)} 
\end{align}
We consider the initial condition $\cU (0) = 1$ and $\Delta (0) = 0$ and the perturbed data are given by $\cU \equiv \cU (1)$ and $\Delta \equiv \Delta (1)$ where $t \in [0 ,1]$ is a real parameter. 
In particular, 
once a perturbation $\Delta $ satisfying $(Q + \Delta )^{2} = 0$ is given, 
one can construct the perturbed data satisfying $(q_{\Delta })^{2} = 0$, 
$\pi _{\Delta } Q_{\Delta } = q_{\Delta } \pi _{\Delta }$ and $\iota _{\Delta } q_{\Delta } = Q_{\Delta } \iota _{\Delta }$ explicitly and obtain the perturbed Hodge type decomposition $1= Q_{\Delta } h_{\Delta } + h_{\Delta } Q_{\Delta } + \iota _{\Delta } \pi _{\Delta }$. 
Formally, 
this correspondence $\cU$ may be invertible because $- \Delta $ also gives a perturbation for $Q_{\Delta }$. 

\vspace{2mm} 

The lemma enables us to construct a morphism $\cU $ concretely, 
which often provides a direct connection or dictionary between two theories. 
As we will see, 
this correspondence $\cU $ induces a similarity transformation such as (\ref{lc of BRST}), 
for which we also write $Q_{\Delta } = \cU ^{-1} Q \, \cU$, 
when a perturbation $\Delta $ does not change the cohomology. 
In section 3 and 4, 
we showed that the light-cone reduction is obtained by applying the lemma (\ref{HPL}). 
In addition to an exact (partial) gauge-fixing which takes into account interactions, 
effective field theory or Wilsonian action may be described by this framework as long as the theory satisfies the (quantum) BV master equation \cite{Doubek:2017naz}.

\subsection{Proof of the lemma}

We give a short proof of the lemma based on formal and algebraic computations. 
For more rigorous treatment, 
consult some mathematical manuscript. 

\vspace{2mm} 

Note that $(Q_{\Delta })^{2} = 0$ holds by construction of $Q_{\Delta } = Q + \Delta$. 
Because of $(Q)^{2} = 0$, we have $(\Delta )^{2} + \Delta Q + Q \Delta = 0$.
We first prove that it implies the relation  
\begin{align*}
A \, \iota \pi  A + A \, Q + Q \, A = 0 \, . 
\end{align*}
One can check it by direct computations:  
\begin{align*}
(l.h.s.) & = 
A (\1 _{\cH } -Q \, h - h \, Q) A + Q A + A Q
= A^{2} + A Q \underbrace{(1- h \, A )}_{= \frac{1}{1 + h \Delta } } 
+ \underbrace{(1 - A \, h )}_{= \frac{1}{1 + \Delta h} } Q A 
\no 
& = \frac{1}{1+\Delta h} \Big[ 
\underbrace{(1+\Delta h) A}_{\Delta } \underbrace{A (1 + h \Delta ) }_{ \Delta }
+ \underbrace{(1+\Delta h) A}_{(A + \Delta h A) = \Delta }  Q 
+ Q \underbrace{A (1 + h \Delta )}_{(A + A h \Delta ) = \Delta }
\Big] \frac{1}{1+h \Delta } 
\no 
& = \frac{1}{1+\Delta h} \Big[ 
(\Delta )^{2} + \Delta Q + Q \Delta 
\Big] \frac{1}{1+h \Delta } 
= 0 \, .
\end{align*}

\subsubsection*{Differential: $(q_{\Delta })^{2} = 0$}

Using $q \pi = \pi Q$, $\iota q = Q \iota $ and $A\iota \pi A + AQ+QA = 0$, we find 
\begin{align*}
(q_{\Delta } )^{2} & = ( q + \pi A \iota )^{2} = (q \pi ) A \iota  + \pi A (\iota q) + \pi ( A \iota \pi A ) \iota 
\no 
& = ( \pi Q) A \iota  + \pi A (Q \iota ) - \pi ( QA + AQ ) \iota = 0 \, . 
\end{align*}

\subsubsection*{Injection: $\iota _{\Delta } q_{\Delta } = Q_{\Delta } \iota _{\Delta }$}

Using $\iota q = Q \iota $, $\iota \pi + Qh + hQ = \1 _{\cH}$, $A \iota \pi A + A Q + Q A = 0$ and $\Delta h A = \Delta - A$, we find 
\begin{align*}
\iota _{\Delta } q_{\Delta } - Q_{\Delta } \iota _{\Delta } 
& = (\iota - h A \iota )(q + \pi A \iota ) - (Q + \Delta ) (\iota - h A \iota ) 
\no 
& = \underbrace{\iota q }_{\alpha } + \iota \pi A \iota 
- hA \underbrace{(\iota q)}_{Q \iota } + h \underbrace{(-A \iota \pi A)}_{Q A + A Q} \iota 
- \underbrace{Q \iota }_{\alpha } + Q h A \iota 
- \Delta \iota + \underbrace{(\Delta h A)}_{\Delta - A} \iota 
\no 
& = ( \iota \pi + h Q  + Q h ) A \iota - A \iota = 0 \, . 
\end{align*}

\subsubsection*{Projection: $q_{\Delta } \pi _{\Delta } = \pi _{\Delta } Q_{\Delta } $}

Using $q \pi = \pi Q$, $\iota \pi + Qh + hQ = \1 _{\cH}$, $A \iota \pi A + AQ +QA =0$ and $A h\Delta = \Delta - A$, we find 
\begin{align*}
q_{\Delta } \pi _{\Delta } - \pi _{\Delta } Q_{\Delta }
& = (q + \pi A \iota )(\pi - \pi A h ) - (\pi - \pi A h )(Q + \Delta )
\no 
& = \underbrace{q \pi }_{\alpha } - \underbrace{(q \pi )}_{\pi Q} A h + \pi A \iota \pi + \pi \underbrace{( - A \iota \pi A )}_{AQ + QA} h 
- \underbrace{\pi Q}_{\alpha } - \pi \Delta + \pi A h Q + \pi \underbrace{(A h \Delta )}_{\Delta - A} 
\no 
& = \pi A (\iota \pi + Q h + h Q ) - \pi A = 0 \, . 
\end{align*}

\subsubsection*{Hodge type decomposition: $Q_{\Delta } h_{\Delta } + h_{\Delta } Q_{\Delta } = \1 _{\cH} - \iota _{\Delta } \pi _{\Delta } $}

Using $A \iota \pi A + AQ+QA=0$, $\Delta h A = A h \Delta = \Delta - A$ and $\iota \pi + Qh + hQ = \1 _{\cH}$, we find 
\begin{align*}
Q_{\Delta } h_{\Delta } + h_{\Delta } Q_{\Delta } + \iota _{\Delta } \pi _{\Delta } & = 
(Q + \Delta )(h - hAh) + (h-hAh)(Q+\Delta ) + (\iota - h A \iota )(\pi - \pi Ah)
\no 
&= \underbrace{Qh}_{\alpha } - QhAh + \Delta h - \underbrace{(\Delta h A)}_{\Delta - A} h 
+ \underbrace{h Q}_{\beta } + h \Delta 
- hAh Q - h \underbrace{(A h \Delta )}_{\Delta -A} 
\no & \hspace{10mm} 
+ \underbrace{\iota \pi }_{\gamma } - \iota \pi A h - hA \iota \pi - h \underbrace{(-A \iota \pi A )}_{AQ+QA} h  
\no 
& = \1_{\cH } - \underbrace{(Qh + hQ + \iota \pi - \1 _{\cH } )}_{\alpha + \beta + \gamma -1 = 0}Ah 
- hA\underbrace{( Q h + hQ + \iota \pi - \1 _{\cH } )}_{\alpha + \beta + \gamma - 1 = 0} \, . 
\end{align*}

One can find that $\pi_{\Delta }$ and $\iota _{\Delta }$ are quasi isomorphisms---morphisms preserving the cohomology, 
for which see mathematical textbooks.

\subsection{A few examples} 

A first example is a trivial situation, 
which is the case of $h = 0$ as follows  
\begin{align*}
h = 0\, 
\rotatebox{-70}{$\circlearrowright $} \, 
\big{(} \cH , Q  \big{)} 
\hspace{3mm} 
\overset{\pi }{\underset{\iota }{
\scalebox{2}[1]{$\rightleftarrows $}
}}
\hspace{3mm} 
\big{(} \cL , q \big{)} 
\hspace{7mm}
\mathrm{with} 
\hspace{5mm} 
 \1 _{\cH } - \iota \, \pi = 0 \, . 
\end{align*}
Then, since perturbed data are given by $\pi _{\Delta } = \pi $, $\iota _{\Delta } = \iota $, $h_{\Delta } = 0$ and $A = \Delta $, we get 
\begin{align*}
h_{\Delta } = 0 \, 
\rotatebox{-70}{$\circlearrowright $} \, 
\big{(} \cH , Q + \Delta \big{)} 
\hspace{3mm} 
\overset{\pi _{\Delta } = \pi }{\underset{\iota _{\Delta } = \iota }{
\scalebox{3.5}[1]{$\rightleftarrows $}
}}
\hspace{3mm} 
\big{(} \cL , q + \pi \Delta \iota \big{)} 
\hspace{7mm}
\mathrm{with} 
\hspace{5mm} 
 \1 _{\cH } - \iota _{\Delta } \, \pi _{\Delta } = 0 \, . 
\end{align*}
In particular, 
it preserves the homotopy equivalence relation $\iota _{\Delta } \, \pi _{\Delta } = 1 = \iota \, \pi = 0$.

\subsubsection*{Contractible situation}

The second example is a contractible situation, 
namely, 
$\iota \, \pi = 0$. 
When a given complex $(\cH _{C} , Q )$ is contractible,   
\begin{align*}
h \, 
\rotatebox{-70}{$\circlearrowright $} \, 
\big{(} \cH _{C} , Q \big{)} 
\hspace{3mm} 
\overset{\pi }{\underset{\iota }{
\scalebox{2}[1]{$\rightleftarrows $}
}}
\hspace{3mm} 
\big{(} \cL = 0 , q \big{)} 
\hspace{7mm}
\mathrm{with} 
\hspace{5mm} 
 \1 _{\cH } = Q \, h + h \, Q  \, , 
\end{align*}
the perturbed one $(\cH _{C} , Q + \Delta )$ is also contractible with $h_{\Delta } = h - h A h$, 
\begin{align*}
h_{\Delta } \, 
\rotatebox{-70}{$\circlearrowright $} \, 
\big{(} \cH _{C} , Q+\Delta \big{)} 
\hspace{3mm} 
\overset{\pi _{\Delta }}{\underset{\iota _{\Delta }}{
\scalebox{2}[1]{$\rightleftarrows $}
}}
\hspace{3mm} 
\big{(} \cL = 0 , q_{\Delta } \big{)} 
\hspace{7mm}
\mathrm{with} 
\hspace{5mm} 
 \1 _{\cH } = Q_{\Delta } \, h_{\Delta } + h_{\Delta } \, Q_{\Delta } \, .  
\end{align*}

Let us consider an example of contractible situation. 
We assume that $1 + \Delta h$ is invertible on $\cH _{n}$ and $\cH _{n+1}$. 
The following sequence gives an example, 
\begin{align*}
\cH _{n-1}
\hspace{3mm} 
\overset{Q}{\underset{h}{
\scalebox{2}[1]{$\rightleftarrows $}
}}
\hspace{3mm} 
\cH _{n}
\hspace{3mm} 
\overset{Q}{\underset{h}{
\scalebox{2}[1]{$\rightleftarrows $}
}}
\hspace{3mm} 
\cH_{n+1}
\hspace{7mm} 
\mathrm{with} 
\hspace{5mm} 
\1 _{\cH _{n}} = Q \, h + h \, Q \, . 
\end{align*}
We find that $h_{\Delta } = h - h A h$ gives a contracting homotopy of $Q_{\Delta } = Q +\Delta $ on $\cH _{n}$. 
Note that $h = Q_{n}^{-1} (1_{\cH _{n}} - Q_{n+1}^{-1} Q) \delta (\cH _{n} )+ Q_{n+1}^{-1} \pi \delta (\cH _{n+1})$ defines the contraction using a left inverse of the inclusion $\iota : \mathrm{Im} (Q) \rightarrow \cH _{n+1}$ and a right inverse $Q^{-1}_{k+1} : \mathrm{Im}(Q) \rightarrow \cH _{k}$ for $k=n-1 , n$.

\subsubsection*{Direct sum of standard and contractible situations} 

Let us consider a complex of the direct sum of $\cH$ and $\cH _{C}$ with $\cD$,  
\begin{align*}
\cdots 
\longrightarrow \, 
\cH \oplus \cH _{C} \, 
\overset{\cD }{
\longrightarrow } \, 
\cH \oplus \cH _{C} \, 
\longrightarrow \, 
\cdots 
\hspace{5mm} 
\mathrm{with} 
\hspace{5mm} 
\cD = 
\begin{bmatrix}
a & b \\ 
c & d 
\end{bmatrix} \, , 
\end{align*}
in which $\cD$ is nilpotent and $( \phi , \lambda ) \in \cH \oplus \cH _{C}$ is mapped by $\cD$ as follows 
\begin{align*}
\cD \, : \,  \phi \oplus \lambda \, \longmapsto \, \big{[} a \phi + b \lambda \big{]} \oplus \big{[}c \phi + d \lambda \big{]} \, . 
\end{align*} 
We assume that $(\cH _{C} , d)$ is contractible with $h$, 
namely, 
$\1 _{\cH _{C} } = d h + h d$. 
Then, 
one can consider the following standard situation  
\begin{align*}
H = 0 \oplus h \, 
\rotatebox{-70}{$\circlearrowright $} \, 
\big{(} \cH \oplus \cH _{C} , 0 \oplus d \big{)} 
\hspace{3mm} 
\overset{\pi }{\underset{\iota }{
\scalebox{2}[1]{$\rightleftarrows $}
}}
\hspace{3mm} 
\big{(} \cL \oplus \underbrace{\cL _{C} }_{0} , 0 \big{)} 
\hspace{7mm}
\mathrm{with} 
\hspace{5mm} 
\alpha \oplus \beta = 
\begin{bmatrix} 
\alpha & 0 \\ 
0 & \beta  
\end{bmatrix} 
\, , 
\end{align*}
where we used trivial injection $\iota $ and projection $\pi : \phi \oplus \lambda \mapsto \phi \oplus 0$. 
The operators 
\begin{align*}
\Delta = 
\begin{bmatrix}
a & b \\
c & 0 
\end{bmatrix} 
\, , \hspace{10mm} 
\cA = \Delta \sum _{n=0}^{\infty } ( -h \Delta )^{n}
= 
\begin{bmatrix}
a + b h c & 0 \\
0 & 0 
\end{bmatrix}  \, , 
\end{align*}
can be used to perturb the above standard situation. 
Since the perturbed data are 
\begin{align*}
\iota _{\Delta }(\phi \oplus 0 ) = \phi \oplus [ h c \phi ] \, , 
\hspace{5mm} 
\pi _{\Delta } ( \phi \oplus \lambda ) = [\phi + b h \lambda ] \oplus 0 \, , 
\hspace{5mm} 
H_{\Delta } = H =  
\begin{bmatrix}
0 & 0 \\ 
0 & h 
\end{bmatrix} \, , 
\end{align*}
we obtain the following standard situation with $\cD = 1 \oplus d + \Delta$ and $\cA = \pi \cA \iota $, 
\begin{align*}
H_{\Delta } = H  \, 
\rotatebox{-70}{$\circlearrowright $} \, 
\big{(} \cH \oplus \cH _{C} , \cD  \big{)} 
\hspace{3mm} 
\overset{\pi }{\underset{\iota }{
\scalebox{2}[1]{$\rightleftarrows $}
}}
\hspace{3mm} 
\big{(} \cL \oplus \underbrace{\cL _{C}}_{0} ,\cA \big{)} 
\hspace{7mm}
\mathrm{with} 
\hspace{5mm} 
\cA = 
\begin{bmatrix}
a + b h c & 0 \\ 
0 & 0 
\end{bmatrix} \, . 
\end{align*}

\subsection{Application to similarity transformations}

We explain that similarity transformations of the BRST operator can be constructed by applying an infinitesimal version of the homological perturbation lemma and by solving the differential equation for a morphism $\cU (t)$ satisfying $\cU (0) =1$. 
Let us consider an infinitesimal perturbation $\Delta (t) \equiv \Delta \, dt $ satisfying $(dt)^{2}=0$ in (\ref{HPL}). 
We find $A = \Delta \, dt$ because of $(dt)^{2} = 0$. 
Likewise, we find that (\ref{injection}) and (\ref{projection}) provide 
\begin{align*}
\boldsymbol{\iota _{\Delta dt} } - \boldsymbol{\iota } =  - ( \boldsymbol{h \Delta } ) \, \boldsymbol{\iota _{\Delta dt }} \, dt \, , 
\hspace{5mm} 
\boldsymbol{\pi  _{\Delta dt}} -  \boldsymbol{\pi } = - \boldsymbol{\pi _{\Delta dt }} ( \boldsymbol{\Delta h} ) \, dt \, .  
\end{align*}
We assume that when the perturbation is infinitesimal, 
the morphism $\cU (t)$ satisfies 
\begin{align}
\label{assumption}
\boldsymbol{\iota _{\Delta dt } } = \boldsymbol{\cU ^{-1} (t) } \, \boldsymbol{\iota } \, , 
\hspace{5mm} 
\boldsymbol{\pi _{\Delta dt} } = \boldsymbol{\pi } \, \boldsymbol{\cU  (t) } \, . 
\end{align}
It provides the relation $\cU (t) (\iota  \pi )_{\Delta dt} = (\iota \pi ) \, \cU (t) $ in (\ref{HPL}). 
By using a formal power series $\cU ( t ) = \cU (0) + \frac{d \, \cU (t) }{dt} \big{|}_{t=0} \, dt + \cdots$, 
we obtain  
\begin{align}
\label{U-eq} 
\bigg{[} 
\Big{(} 
\boldsymbol{ \frac{d}{dt} } +  \boldsymbol{h \, \Delta }  
\Big{)} \, 
\boldsymbol{\cU ^{-1} (t) } \, 
\bigg{]} \, 
\boldsymbol{\iota } = 0 \, , 
\hspace{7mm} 
\boldsymbol{\pi } \, \bigg{[}  
\boldsymbol{ \frac{d \, \cU (t) }{dt} } 
+ \boldsymbol{\cU (t)}  \boldsymbol{\Delta \, h }  
\bigg{]} = 0 \, , 
\end{align} 
which determines $\cU (t)$ up to operating $\iota $ or $\pi $. 
Note that $\pi \, \cU \, ( \Delta h + h \Delta ) \, \cU ^{-1} \iota = 0$. 
In order to find a similarity transformation, 
it may be helpful to consider a situation that $\pi $ or $\iota $ can be identified with $\cU (t )$ in (\ref{HPL}), 
which is consistent with our assumption (\ref{assumption}). 

\vspace{2mm} 

Let us consider the transformation $\cU _{1} (t)$ connecting $c_{0} K^{lc} + d$ to $Q - \sum_{k}Q_{k}$, 
\begin{align*}
c_{0} \big{[} K^{lc} + t \, N \big{]} + d = \cU ^{-1}_{1} (t) \big{(} c_{0} K^{lc} + d \big{)} \, \cU _{1}(t) \, , 
\end{align*}
where $\cU _{1} (0) =1$ and thus $t \, \Delta = t \, c_{0} N$ can be regarded as an infinitesimal perturbation. 
We use (\ref{lc h}) as a homotopy contracting operator because $\Delta = c_{0} N$ vanishes on $\Pi $ of (\ref{lc h}). 
In this case, 
$\cU _{1} (t)$ preserves the homotopy equivalence relation $\iota \, \pi + Q \, h + h \, Q  = 1 = \iota _{\Delta } \pi _{\Delta } + Q_{\Delta } \, h_{\Delta } + h_{\Delta } \, Q_{\Delta }$ because $\Delta $ commutes with $h$ and $\Pi = \iota \, \pi$. 
By substituting $\Delta = c_{0} N$ into (\ref{U-eq}), 
we find that a defining equation of $\cU _{1} $ is given by  
\begin{align}
\label{def of U1}
\frac{d}{dt} \cU _{1} (t) = \big{(} h \, c_{0} N \big{)} \, \cU _{1} (t) 
= - c_{0} \kappa \, \, \cU _{1} (t) \, .
\end{align}
We obtain a solution $\cU _{1} (t) = e^{- t c_{0} \kappa }$ with the initial condition $\cU _{1} (0) = 1$, 
which gives the similarity transformation (\ref{U1}) at $t =1$. 

\vspace{2mm}

Likewise, we can get (\ref{U2}) by using the infinitesimal perturbation. 
Let us consider the infinitesimal transformation $\cU _{2} (t)$ satisfying $\cU _{2} (0) = 1$ and  
\begin{align*}
d + c_{0} [K^{lc} + N ] + dt \, \sum_{k=1}^{3} Q_{k}  
= \cU ^{-1}_{2}(t) \big{(} d + c_{0} [K^{lc} + N ] \big{)} \, \cU _{2} (t) \, . 
\end{align*}
We would like to use $dt \sum_{k} Q_{k}$ as an infinitesimal perturbation. 
Unlike $c_{0} N$, 
the operator $\sum _{k} Q_{k}$ itself does not give an infinitesimal perturbation without $(dt)^{2} = 0$. 
We write $\Pi _{\pm }$ for the projector onto the kernel of $S_{\pm }$. 
Since $\ld S_{\pm } , Q_{k} \rd \not= 0$, 
one may regard $\Pi _{\pm } = \iota \, \pi $ in (\ref{HPL}) in order to apply (\ref{assumption}). 
Because of $\ld d , \kappa _{\pm } \rd = S_{\pm }$, 
we find a homotopy contracting operator $h_{\pm }$ satisfying $\ld d , h_{\pm } \rd = 1 - \Pi _{\pm }$ is given by 
\begin{align*}
h_{\pm } \equiv S^{-1}_{\pm } \kappa _{\pm } \, , 
\hspace{8mm} 
S^{-1}_{\pm } \equiv \frac{1}{S_{\pm } } \big{(} 1 - \Pi _{\pm } \big{)} \, . 
\end{align*}
We use this $h_{\pm }$ as a homotopy contracting operator. 
In this case, 
$\cU _{2} (t)$ changes a homotopy equivalence relation from the initial one $\Pi _{\pm } + \ld d , h_{\pm } \rd = 1$ because of $\Delta \, h_{\pm } + h_{\pm } \, \Delta \not= 0$. 
Note that $\ld S^{-1}_{\pm } , \kappa _{\pm } \rd = 0$, 
$S^{-1}_{\pm } \Pi _{\pm } = \frac{1}{S_{\pm } } ( 1 - \Pi _{\pm } ) \Pi _{\pm } = 0$ and $S_{\pm } S_{\pm}^{-1}  = S^{-1}_{\pm } S_{\pm }= (1-\Pi _{\pm })$. 
However, 
unlike (\ref{def of U1}), 
a naive equation obtained by substituting $\Delta dt = dt \sum_{k} Q_{k}$ into (\ref{U-eq}) is not consistent if $\iota $ or $\pi $ is removed. 
We find that at $t=0$, 
the term 
\begin{align*} 
(h_{\pm } \Delta ) \, \iota \, \pi = \kappa _{\pm } (Q_{1} + \frac{1}{2} Q_{2} + \frac{1}{2} Q_{3} ) \, \iota \, \pi 
\end{align*} 
appears in (\ref{U-eq}) because $\ld S_{\pm }^{-1} , Q_{1} \rd = Q_{1}$, 
$\ld S_{\pm }^{-1} , Q_{2} \rd = \frac{1}{2} Q_{2}$ and $\ld S_{\pm }^{-1} , Q_{3} \rd = \frac{1}{2} Q_{3}$ hold on $\Pi _{\pm } = \iota \pi $, 
for which we write $\ld S^{-1}_{\pm } , Q_{k} \rd _{\Pi _{\pm } }$. 
A symmetrized equation compatible with $(\frac{d}{dt} \cU ) \, \cU ^{-1}+ \cU \, ( \frac{d}{dt} \cU ^{-1}) = 0$ is given by 
\begin{align*}
\Big{(} 
\frac{d}{dt} \cU _{2} (t) 
\Big{)} \cU ^{-1}_{2} (t) 
& = \cU _{2} (t) \Big{[} \kappa _{\pm }  \Ld S^{-1}_{\pm } , \sum_{k=1}^{3} Q_{k} \Rd + \Ld S_{\pm }^{-1} , \sum_{k=1}^{3} Q_{k} \Rd \, \kappa _{\pm } \Big{]}_{\Pi _{\pm }} \, \cU _{2}^{-1} (t) \, . 
\end{align*}
Therefore, 
by using $\ld \kappa _{\pm } , Q_{k} \rd = k\, R_{k+1}$ for $k=1,2$ and $\ld \kappa _{\pm } , Q_{3} \rd = 0$, 
we obtain a defining equation of $\cU _{2}$ as follows 
\begin{align} 
\frac{d}{dt} \, \cU _{2} (t) = \cU _{2} (t) \big{(} R_{2} + R_{3} \big{)} \,  . 
\end{align}
The initial condition $\cU _{2} (0) = 1$ gives a simple solution $\cU _{2} (t) = e^{t (R_{2} + R_{3} )}$, 
which gives the similarity transformation (\ref{U2}) at $t=1$.


{\footnotesize 

\begin{thebibliography}{99}

\bibitem{Witten:1985cc}
  E.~Witten,
  ``Noncommutative Geometry and String Field Theory,''
  Nucl.\ Phys.\  B {\bf 268}, 253 (1986).
\bibitem{Zwiebach:1992ie}
  B.~Zwiebach,
  ``Closed string field theory: Quantum action and the B-V master equation,''
  Nucl.\ Phys.\  B {\bf 390}, 33 (1993)
  [arXiv:hep-th/9206084].

  
\bibitem{Kaku:1974zz}
  M.~Kaku and K.~Kikkawa,
  ``The Field Theory of Relativistic Strings, Pt. 1. Trees,''
  Phys.\ Rev.\ D {\bf 10} (1974) 1110. 
  M.~Kaku and K.~Kikkawa,
  ``The Field Theory of Relativistic Strings. 2. Loops and Pomerons,''
  Phys.\ Rev.\ D {\bf 10} (1974) 1823.


\bibitem{Hata:1986jd}
  H.~Hata, K.~Itoh, T.~Kugo, H.~Kunitomo and K.~Ogawa,
  ``Covariant String Field Theory,''
  Phys.\ Rev.\ D {\bf 34} (1986) 2360.
  H.~Hata, K.~Itoh, T.~Kugo, H.~Kunitomo and K.~Ogawa,
  ``Covariant String Field Theory. 2.,''
  Phys.\ Rev.\ D {\bf 35} (1987) 1318.
\bibitem{Kugo:1987rq}
  T.~Kugo,
  ``Covariantized Light Cone String Field Theory,''
  In *Santiago 1987, Proceedings, Quantum mechanics of fundamental systems 2* 167-187 and Kyoto Univ. - KUNS-0917 (88,rec.Apr.) 36 p
\bibitem{Siegel:1987ku}
  W.~Siegel and B.~Zwiebach,
  ``Interacting {BRST} From the Light Cone,''
  Nucl.\ Phys.\ B {\bf 299} (1988) 206.


\bibitem{Kato:1982im}
  M.~Kato and K.~Ogawa,
  ``Covariant Quantization of String Based on BRS Invariance,''
  Nucl.\ Phys.\ B {\bf 212} (1983) 443.
\bibitem{Aisaka:2004ga}
  Y.~Aisaka and Y.~Kazama,
  ``Relating Green-Schwarz and extended pure spinor formalisms by similarity transformation,''
  JHEP {\bf 0404} (2004) 070
  [hep-th/0404141].


\bibitem{Crainic}
  M.~Crainic, 
  ``On the perturbation lemma, and deformations,'' 
  arXiv:math/0403266 [math.AT]. 
\bibitem{Berglund}
  A.~Berglund, 
  ``Homological perturbation lemma for algebra over operads,'' 
  Algebr. Geom. Topol. 14 (2014) 2511-2548 
  [arXiv::0909.3485 [math.AT]]. 
\bibitem{Vallette} 
  B.~Valltte, 
  ``Algebra + Homotopy = Operad,'' 
  arXiv:1202.3245 [math.AT]. 


\bibitem{Henneaux} 
  M.~Henneaux and C.~Teitelboim, 
  ``Quantization of gauge systems,'' 
  (Princeton University Press), 1992. 
\bibitem{Aisaka:2002sd}
  Y.~Aisaka and Y.~Kazama, 
  ``A New first class algebra, homological perturbation and extension of pure spinor formalism for superstring,'' 
  JHEP {\bf 0302} (2003) 017 
  [hep-th/0212316]. 


\bibitem{Kajiura:2003ax}
  H.~Kajiura,
  ``Noncommutative homotopy algebras associated with open strings,''
  Rev.\ Math.\ Phys.\  {\bf 19} (2007) 1.
  [math/0306332 [math-qa]].
  H.~Kajiura,
  ``Homotopy algebra morphism and geometry of classical string field theory,''
  Nucl.\ Phys.\ B {\bf 630} (2002) 361.
  [hep-th/0112228]. 
\bibitem{Konopka:2015tta}
  S.~Konopka,
  ``The S-Matrix of superstring field theory,''
  JHEP {\bf 1511} (2015) 187
  [arXiv:1507.08250 [hep-th]]. 
\bibitem{Erler:2016rxg}
  T.~Erler,
  ``Supersymmetry in Open Superstring Field Theory,''
  JHEP {\bf 1705} (2017) 113
  [arXiv:1610.03251 [hep-th]].


\bibitem{Doubek:2017naz}
  M.~Doubek, B.~Jurco and J.~Pulmann,
  ``Quantum $L_{\infty }$ Algebras and the Homological Perturbation Lemma,''
  arXiv:1712.02696 [math-ph].


\bibitem{Matsunaga:2016}
  H.~Matsunaga,
  ``Notes on the Wess-Zumino-Witten-like structure: $L_{\infty }$ triplet and NS-NS superstring field theory,''
  JHEP {\bf 1705} (2017) 095
  [arXiv:1612.08827 [hep-th]].
\bibitem{Erler:2017onq}
  T.~Erler,
  ``Superstring Field Theory and the Wess-Zumino-Witten Action,''
  JHEP {\bf 1710} (2017) 057
  [arXiv:1706.02629 [hep-th]].


\bibitem{Sen:2016qap}
  A.~Sen,
  ``Wilsonian Effective Action of Superstring Theory,''
  JHEP {\bf 1701} (2017) 108
  [arXiv:1609.00459 [hep-th]].

\end{thebibliography}
}
\end{document}